\documentclass[fleqn,usenatbib]{mnras}
\usepackage{newtxtext,newtxmath}
\usepackage{comment}
\usepackage[T1]{fontenc}
\usepackage{placeins}
\usepackage{float}

\DeclareRobustCommand{\VAN}[3]{#2}
\let\VANthebibliography\thebibliography
\def\thebibliography{\DeclareRobustCommand{\VAN}[3]{##3}\VANthebibliography}

\usepackage{graphicx}	
\usepackage{amsmath}	

\title[Statistical imprints of wave-like dark matter in JWST]
{Statistical imprints of wave-like dark matter on multiply-imaged galaxies in strong cluster lenses from JWST}

\author[Ephremidze et al.]{
Nino Ephremidze,$^{1}$\thanks{E-mail: nino\_ephremidze@g.harvard.edu}
Daniel Gilman$^{2}$\thanks{E-mail:
gilmanda@uchicago.edu},
and Cora Dvorkin$^{1}$\thanks{E-mail: cdvorkin@g.harvard.edu}
\\
$^{1}$Department of Physics, Harvard University, Cambridge, MA 02138, USA\\
$^{2}$Department of Astronomy \& Astrophysics, University of Chicago, Chicago, IL 60637, USA
}

\pubyear{\the\year{}}

\begin{document}
\label{firstpage}
\pagerange{\pageref{firstpage}--\pageref{lastpage}}
\maketitle

\begin{abstract}
Wave-like dark matter ($\psi$DM) is an elusive dark matter (DM) candidate. The model, often also called fuzzy or ultralight DM, proposes that DM is an extremely light ($m\sim10^{-22}$ eV) boson and thereby has a kpc-scale de Broglie wavelength. Hence, interference of DM gives rise to sub-galactic density fluctuations that can be studied with strong gravitational lensing. In this paper, we use the residual power spectrum, $\mathrm{P}_{\delta}(k)$, as a probe of $\psi$DM, which quantifies deviations from smooth lensing predictions, measured from multiply-imaged galaxies in strong cluster lenses. The key idea is that imprinted in these deviations are lensing distortions from DM substructure, which can be harnessed statistically to distinguish among DM theories. We simulate JWST-quality mock observations of strong gravitational lensing in galaxy clusters, modeling line-of-sight DM substructure within $\psi$DM and the standard cold dark matter (CDM) paradigms. Using mock deep observations ($\sim$ 20 hours), we find that $\mathrm{P}_{\delta}(k)$ is sensitive to both $\psi$DM particle mass and fluctuation amplitude, and can distinguish $\psi$DM fluctuations from CDM subhalos. We demonstrate that $\mathrm{P}_{\delta}(k)$ can be measured directly from data by modeling the smooth lensing with a local Curved Arc Basis formalism. With realistic modeling systematics, we find a statistically significant separation between $\psi$DM and CDM across $1 \lesssim k \lesssim 11\,\mathrm{kpc}^{-1}$ -- offering an independent probe of the wave-like nature of DM complementary to existing constraints.
\end{abstract}

\begin{keywords}
gravitational lensing: strong -- dark matter -- galaxies: clusters
\end{keywords}

\section{Introduction}
\label{sec:intro}
 
The existence of dark matter (DM) is supported by overwhelming evidence spanning
galactic to cosmological scales, from the rotation curves of galaxies to the
large-scale structure of the universe and the cosmic microwave background
\citep[CMB;][]{Bennett2003,Planck2020,Louis2025}. The standard cosmological model
treats DM as a collisionless, pressureless fluid -- cold dark matter (CDM)
-- and this paradigm successfully reproduces a vast array of observations at
large scales \citep{Springel2005,Gil-marin2016,Abbott2018,Abolfathi2018,DESI2024iii}.  The leading CDM particle candidate is the weakly
interacting massive particle (WIMP).  However, extensive direct-detection
experiments have returned null results and have excluded large fractions of the
simplest WIMP parameter space \citep{Aalbers2025,Aprile2025,Bo2025}.  At the
same time, the CDM model faces persistent challenges at sub-galactic scales,
including the cusp–core problem \citep{Moore1994,Moore1999a,Flores1994,Oh2015,
Gentile2004}, the missing-satellite problem \citep{Moore1999b,Klypin1999}, and
the too-big-to-fail problem \citep{Boylan2011,Boylan2012}.  Although baryonic
feedback processes, such as supernova-driven outflows, can alleviate some of these
tensions, it remains unclear whether such mechanisms can fully resolve all
small-scale discrepancies \citep{Weinberg2015,Schaller2015}.
 
These difficulties, together with the non-detection of WIMP candidates in the
laboratory, have motivated the development of alternative DM theories.  To be
compelling, such alternatives should be physically motivated, reproduce CDM
predictions on large scales, yet leave measurable astrophysical signatures that
distinguish them from CDM.  Wave-like dark matter -- also called fuzzy dark
matter (FDM) or ultralight dark matter (ULDM); we use these terms interchangeably and adopt the notation $\psi$DM throughout -- has emerged as one of the
most actively studied CDM alternatives satisfying all three criteria.  In this
framework, the DM particle is an extraordinarily light ($m_\psi \sim
10^{-22}$~eV) boson, whose associated de~Broglie wavelength reaches
kiloparsec scales inside galactic halos.  As a consequence, wave interference
of the macroscopic DM wave function produces order-unity density fluctuations on
the de~Broglie scale, suppresses the formation of structure below the Jeans
scale, and generates solitonic cores at halo centers -- all in a regime that is
inaccessible to standard CDM simulations \citep{Hu2000,Schive2014a,
Hui2017,Hui2021}.  Particle-physics motivations for such ultralight bosons arise
naturally in string-theory axiverse scenarios and
axion-like particle (ALP) models \citep{Marsh2016}.

Observational constraints on the $\psi$DM particle mass span a wide range of
independent probes. Cosmological observations constrain $\psi$DM through CMB
anisotropies and large-scale structure \citep{Hlozek2015,Lague2022}, the
Lyman-$\alpha$ forest flux power spectrum \citep{Irsic2017,RogersPeiris2021},
and high-redshift galaxy UV luminosity functions from HST and JWST
\citep{Winch2024}. Galactic dynamics provides complementary constraints from
the sizes and stellar kinematics of dwarf galaxies
\citep{Dalal2022,Pozo2024}, subhalo-induced gaps in Milky Way stellar streams
\citep{Schutz2020,Banik2021}, pulsar timing arrays probing the
oscillating gravitational potential of $\psi$DM
\citep{Porayko2018,Smarra2023}, and black-hole superradiance 
\citep{Arvanitaki2015,StottMarsh2018,Hoof2024}.

Strong gravitational lensing has become a particularly powerful probe at the
kpc scales relevant to the canonical $\psi$DM mass. \citet{Chan2020}
demonstrated through simulations that wave-interference fluctuations in
$\psi$DM halos themselves produce flux-ratio anomalies similar to those
attributed to CDM subhalos. Building on this, \citet{Laroche2022}
analyzed flux ratios in eleven quadruply imaged quasars and showed that 
$m_\psi < 10^{-21.5}$ eV is disfavored by data, but not completely ruled out, 
as wave interference fluctuations in the host halo masquerade as CDM subhalos.
\citet{Powell2023} used a single galaxy-scale
lens observed at milli-arcsecond resolution with very long baseline
interferometry (VLBI) to directly constrain FDM granule structures in the
main halo. More recently, \citet{Amruth2023} showed that $\psi$DM lens models
correctly predict the level of brightness and position anomalies in a
quadruply lensed system, whereas standard CDM lens models often
fail, although \citet{MillerWilliams2025} have since argued that the observed
anomalies can be explained by CDM mass models with added mass complexity in
multipoles.
In a cluster-lensing context most relevant to this work,
\citet{Broadhurst2024} used JWST observations of microlensing events in the
Dragon Arc behind Abell~370 to argue that their skewed spatial distribution
favors $\psi$DM with $m_\psi \simeq 10^{-22}$~eV over smooth or CDM-subhalo
predictions.

Several other recent studies have explored FDM signatures in the magnification
properties near radial critical curves \citep{Palencia2025}, wave-optics
phenomena in gravitational-wave lensing \citep{Singh2025}, and time-varying
gravitational potentials measurable with lensed fast radio bursts
\citep{Gao2025}.
 
Despite these advances, with the exception of \citealt{Laroche2022}, current 
constraints on $m_\psi$ are derived from
individual systems and may therefore be vulnerable to poor modeling of
baryonic physics \citep{Hui2021} or to the absence of full wave-interference simulations
that capture higher-order nonlinear effects \citep{Dalal2022}. Accounting for
these factors can potentially relax existing bounds \citep{Palencia2025}.  It is
therefore important to develop novel, independent probes of $m_\psi$ that are
immune to the systematic uncertainties of existing methods, so that constraints
from complementary approaches can be cross-checked.
 
A promising and largely unexplored avenue is to use galaxy clusters -- the most
powerful gravitational lenses in the universe -- as probes of sub-galactic DM
structure.  Deep James Webb Space Telescope (JWST) imaging of cluster lenses
resolves hundreds of multiply imaged background galaxies (e.g., \citealt{Bergamini2023,Rihtarsic2025}), each
carrying a statistical imprint of the DM density along the line of sight.
Global parametric lens modeling of cluster-scale mass distributions is
notoriously complex, but a local lensing formalism -- Curved Arc Basis
\citep[CAB;][]{Birrer2021} -- circumvents this challenge by describing the
gravitational distortions in the immediate vicinity of each arc.
\citet{Sengul2023} demonstrated with JWST-quality simulations of the cluster
lens SMACS~0723 that the CAB approach can detect line-of-sight perturbers 
with masses down to $\sim10^8\,\mathrm{M}_\odot$.

Beyond the detection of individual massive 
subhalos, one can study a statistical, population-level 
signature of $\psi$DM substructure in galaxy clusters.
We use the \emph{residual power spectrum}
$\mathrm{P}_\delta(k)$ \citep{Cyr-Racine:2018htu}, which quantifies 
the statistical deviations of the observed lensed surface brightness 
from the prediction of a smooth lens model.
The key insight is that the amplitude and shape of $\mathrm{P}_\delta(k)$
encode the population statistics of all DM substructure along the line of sight,
allowing $\psi$DM and CDM to be differentiated without resolving individual
subhalos.
A similar probe called the substructure power spectrum (the power spectrum 
of the projected dark matter substructure field), was first proposed as a 
lensing observable by \citet{Hezaveh2014}. It was subsequently developed analytically 
and numerically for CDM, WDM, and self-interacting dark matter
\citep{Rivero2018a,Rivero2018b,Brennan2018} and extended to incorporate
line-of-sight halos and anisotropic subhalo distributions
\citep{Sengul2020,Dhanasingham2023a,Dhanasingham2023b,Dhanasingham2025}.
$\mathrm{P}_\delta(k)$ is the observational counterpart of
this theoretical quantity, which depends on lens configurations, source morphology, and instrumental effects, but is directly measurable from strong-lensing data. 
$\mathrm{P}_\delta(k)$ has been used to constrain sub-galactic mass 
structure with galaxy-galaxy lenses \citep{Bayer2023nwm, Bayer2023jkf}, but has not been 
applied to strong cluster lenses.

In this work, we use the statistical approach of residual power spectra to 
study the characteristic wave-interference signatures of $\psi$DM in 
multiply-imaged galaxies from JWST clusters. Strong cluster lenses provide an 
especially well-suited avenue for this measurement for several reasons. Firstly, 
highly magnified arcs stretch across many JWST pixels, dramatically increasing 
both the flux and the spatial area available for statistical analysis. Secondly, 
the large number of multiple images in cluster lenses breaks degeneracies between 
substructure signals and source modeling flexibility. Finally, the cluster 
lensing geometry probes long lines of sight, accumulating stronger substructure 
signal from line-of-sight halos.

We simulate JWST-quality mock observations of strong gravitational lensing in
galaxy clusters under both the $\psi$DM and CDM paradigms, modeling line-of-sight DM
substructure with \texttt{pyHalo}\footnote{\url{https://github.com/dangilman/pyHalo}}.  Using mock deep exposures
($\sim20$ hours), we demonstrate that $\mathrm{P}_\delta(k)$ is sensitive to
both the $\psi$DM particle mass $m_\psi$ and the fluctuation amplitude, and can
discriminate $\psi$DM fluctuations from CDM subhalos.  We further show that
$\mathrm{P}_\delta(k)$ can be measured directly from data using the CAB local
lens modeling formalism, and that -- with realistic modeling systematics --
the 16th--84th percentile bands of $\psi$DM and CDM are non-overlapping across $1 \lesssim k \lesssim 11\,\mathrm{kpc}^{-1}$, demonstrating a statistically significant separation between the two models.

The paper is organized as follows.
Section~\ref{sec:methods} describes the simulation and analysis pipeline, including the source galaxy, lens cluster, and substructure models used to produce JWST-quality mock observations, and the source and lens modeling techniques used to measure $P_\delta(k)$ in realistic strong-lensing analysis.
We present our results in Section~\ref{sec:results}, discuss their interpretations in Section~\ref{sec:discussion}, and summarize our conclusions in Section~\ref{sec:conclusions}.
  
\section{Methods}
\label{sec:methods}

We simulate JWST-quality mock observations of galaxy sources strongly lensed by a cluster and perturbed by a line-of-sight population of substructure, generated under cold dark matter (CDM) and wave dark matter ($\psi$DM) paradigms. We propose that statistical perturbations from the population of low-mass dark matter structures will be imprinted on the residuals between the observed images and our best-fit image reconstructions with a smooth model. We introduce the residual power spectrum, $P_{\delta}(k)$, as a summary statistic that captures this population-level lensing effect to distinguish the $\psi$DM model from the fiducial CDM scenario. We perform a realistic strong lensing analysis, jointly fitting the source light and smooth component of the lens mass distribution, with a local Curved Arc Basis (\citealt{Birrer2021}) formalism to measure $P_{\delta}(k)$ directly from the data. 

To generate mock observations, we use real galaxy images from the COSMOS catalog (\citealt{Scoville2007}, \citealt{Koekemoer2007}) as light sources, and model the smooth component of the cluster lens as an elliptical Navarro-Frenk-White profile (\citealt{NFW1997}). We simulate a population of CDM or $\psi$DM perturbers within and along the line of sight of the galaxy cluster using \texttt{pyHalo} (\citealt{Gilman2020}). For all gravitational lensing computations, we use \texttt{lenstronomy} (\citealt{Birrer2018}, \citealt{Birrer2021_lenstronomy}), a strong-lensing analysis and simulation software, and we make a new subpackage publicly available for rendering James Webb Space Telescope (JWST) quality mock observations\footnote{\url{https://github.com/ninoephremidze/statistically_probing_subgalactic_dark_matter/JWST.py}}.

\subsection{Galaxy sources}
\label{sec:cosmos_sources}

The light distribution of galaxies can be approximated by analytic light profiles (\citealt{Sersic1963}). However, real galaxies can have complex morphologies, ranging from spiral arms to irregular shapes. This variability in source light distribution will imprint a broader range of features in the lensed images. As modeling systematics prevent us from perfectly fitting the source galaxy or the lens mass distribution, some residuals of these morphological features may remain after strong lensing analysis, contributing to a larger variance in the summary statistic $\mathrm{P}_{\delta}(k)$ for a given dark matter model. Hence, to get realistic estimates of the error bars, we should fully account for the variability in the source galaxies.

To this end, we use real galaxy images taken by the Hubble Space Telescope (HST) in the Cosmic Evolution Survey (COSMOS) (\citealt{Scoville2007}, \citealt{Koekemoer2007}) as source light in our simulations.
We select sources from the COSMOS catalog used in the GREAT3 gravitational lensing challenge (\citealt{Mandelbaum2012}, \citeyear{Mandelbaum2014}), accessed via the \texttt{paltas} package\footnote{\url{https://github.com/swagnercarena/paltas}}.
We filter the catalog to include only galaxies at $z_s \leq 1.0$ (matching our simulation source redshift), with a minimum cutout size of 120 pixels, a minimum flux-radius of 40 pixels, and an apparent AB magnitude brighter than 20, retaining a sample of well-resolved, bright galaxies suitable for gravitational lensing simulations.

We note that the COSMOS images were taken in the F814W band of the HST Advanced Camera for Surveys (ACS), which differs from the JWST NIRCam F150W observations we aim to simulate in two important respects: HST ACS has a coarser angular resolution ($0.05''$\,pixel$^{-1}$ native plate scale) and higher pixel noise.
To mitigate these discrepancies, we upsample each HST image to the target JWST pixel scale ($0.031''$\,pixel$^{-1}$) and then convolve the resulting surface brightness map with a Gaussian smoothing kernel of width $\sigma = 0.1''$ to suppress HST pixel noise before lensing.
This ensures that detector noise from HST is not propagated into the lensed images, where it could produce spurious small-scale residuals that mimic the signal of dark matter substructure. The F814W band also samples bluer rest-frame emission than F150W, so typical galaxies would be somewhat brighter in F150W. Hence, using the F814W fluxes makes our signal-to-noise estimates conservative.

To account for source-morphology variability, we draw $N_\mathrm{src} = 20$ independent, randomly-selected COSMOS sources from the filtered catalog for each dark matter model realization, yielding $N_\mathrm{maps} = 60$ residual maps per model (20 sources $\times$ 3 images).
We find that the 16th--84th percentile bands on $P_\delta(k)$ are stable under resampling of source subsets, indicating that 20 realizations are sufficient to characterize the source-morphology variance at the signal-to-noise levels considered here; see Section~\ref{sec:results} for the resulting statistical uncertainties.

\subsection{Cluster lens model}

Galaxy clusters are the largest gravitationally bound objects in the universe, acting as the most powerful gravitational lenses. Similarly to other dark matter halos, their smooth mass distribution is well described by a Navarro-Frenk-White (NFW) profile (\citealt{NFW1997}), given by
\begin{align}
\label{eq:NFW}
    \rho_{\mathrm{NFW}}(r)=\frac{\rho_{\mathrm{s}}}{\frac{r}{r_{\mathrm{s}}}\left(1+\frac{r}{r_{\mathrm{s}}}\right)^2},
\end{align}
where $r_s$ is the scale radius and $\rho_s$ is a characteristic density. The NFW profile can be equivalently parametrized by the virial mass $M_{200}$ and concentration $c_{200} \equiv R_{200}/r_s$, where $R_{200}$ is the radius enclosing a mean density 200 times the critical density. Weak-lensing observations show that cluster mass distributions are moderately elliptical (\citealt{Shin2018}). We therefore model the cluster as an elliptical NFW profile by replacing $r$ in Eq.~\ref{eq:NFW} with the elliptical coordinate $\xi = \sqrt{qx^2 + y^2/q}$, where $q$ is the projected axis ratio (\citealt{Oguri2021}), implemented via the convergent-series expansion (\texttt{NFW\_ELLIPSE\_CSE}) in \texttt{lenstronomy}.

We list all lens model parameters in Table~\ref{tab:cluster_nfw_parameters}. We adopt $M_{200} = 10^{15}\,M_\odot$, placing our simulated cluster among the most massive known systems. This maximizes realistic lensing magnifications and ensures a large caustic area over which multiple-image configurations can be sampled. The concentration $c_{200} = 4$ is derived from the CDM mass-concentration relation of \citealt{Zhao2009} at $z = 0.4$ and is consistent with $\psi$DM predictions on cluster scales (\citealt{Kawai2024}). The lens and source redshifts are $z_\mathrm{lens} = 0.4$ and $z_\mathrm{source} = 1.0$, respectively, chosen to place the system in a typical strong-lensing geometry.

For the ellipticity, we adopt $|e| = 0.271$ from the weak-lensing measurement of \citealt{Shin2018}, where $|e| = (1-q)/(1+q)$ is the reduced ellipticity magnitude, corresponding to an axis ratio $q \approx 0.57$.
In \texttt{lenstronomy}'s complex ellipticity parametrization, this is set as $(e_1, e_2) = (0.27,\, 0.0)$, orienting the cluster major axis along the $x$-direction.

\begin{figure*}
    \centering
    \includegraphics[width=\linewidth]{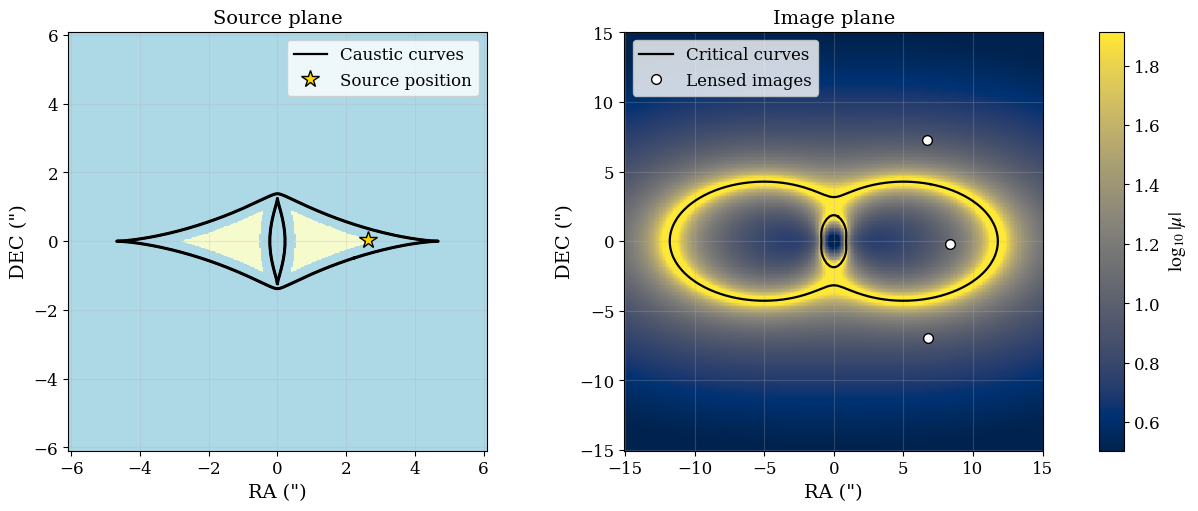}
    \caption{Right: Source plane. The position of a randomly placed source galaxy, resulting in three image locations on the lens plane, is shown as a star. The black solid lines indicate caustic curves, and the shaded yellow region indicates the ``safe" region 0.2 away from the caustic curves. Left: Magnification map on the image plane. The black solid lines show the critical curves, and the three white dots show the positions of the multiply-lensed source galaxy.}
    \label{fig:lens_model}
\end{figure*}

To produce strongly lensed images, we randomly place a given COSMOS source galaxy (see Section~\ref{sec:cosmos_sources}) inside the outer caustic, at a distance $d_{\mathrm{min}} = 0.2''$ from any caustic curve. This lower bound ensures that the critical-curve images are not too closely merged, which would violate the local-approximation assumption of the CAB formalism used in our lens modeling (Section~\ref{sec:methods}). Sources anywhere inside the caustic that satisfy the minimum separation are accepted, yielding three-image configurations across a range of magnifications. We visualize the lens magnification map and critical curves, along with the corresponding caustic curves, for the accessible source region on the source plane in Figure~\ref{fig:lens_model}.

\begin{table}
    \centering
    \begin{tabular}{llll}
        \hline
        Parameter & Symbol & Value & Reference \\
        \hline
        Virial mass          & $M_{200}$   & $10^{15}\,M_\odot$  & --- \\
        Concentration        & $c_{200}$   & $4$                  & \citealt{Zhao2009} \\
        Lens redshift        & $z_\mathrm{lens}$   & $0.4$         & --- \\
        Source redshift      & $z_\mathrm{source}$ & $1.0$         & --- \\
        Ellipticity (aligned) & $e_1$      & $0.27$               & \citealt{Shin2018} \\
        Ellipticity (cross)  & $e_2$       & $0.0$                & --- \\
        Lens centre          & $(x_0, y_0)$ & $(0'',\,0'')$       & --- \\
        Density profile      & ---         & Elliptical NFW    & \citealt{Oguri2021} \\
        \hline
    \end{tabular}
    \caption{Fiducial elliptical NFW cluster lens model parameters. The ellipticity components $e_1, e_2$ follow the lenstronomy reduced complex ellipticity convention, where $|e| = (1-q)/(1+q)$ with $q = b/a$ the projected axis ratio; $e_2 = 0$ aligns the major axis with the $x$-axis. The scale radius $r_s$ and normalization $\alpha_{R_s}$ are derived from $M_{200}$ and $c_{200}$ via \texttt{lenstronomy}'s \texttt{LensCosmo} module assuming Planck cosmology (\citealt{Planck2020}).}
    \label{tab:cluster_nfw_parameters}
\end{table}

\subsection{Dark matter substructure}
\label{sec:dm_substructure}

We simulate dark matter substructure using \texttt{pyHalo} (\citealt{Gilman2020}), a Python package for generating self-consistent realizations of dark matter subhalo populations, line-of-sight (LOS) halos, and associated sub-galactic structures under a specified dark matter model.
We include subhalos up to a maximum infall mass of $M_\mathrm{max} = 10^9\,M_\odot$. More massive objects produce detectable individual perturbations amenable to direct detection methods (\citealt{Sengul2023}) and are excluded from our statistical treatment.
To preserve the mean surface mass density of the lens, we subtract a uniform negative convergence sheet whose amplitude equals the total projected convergence of all rendered substructure.

\subsubsection{Cold Dark Matter}

For CDM, we use pyHalo's \texttt{CDM} class to draw a realization of the subhalo and LOS halo populations.
LOS halos are distributed according to the Sheth--Tormen halo mass function (\citealt{ShethTormen1999}), rescaled by an environment-dependent factor $\xi(M_\mathrm{host}, z)$ that accounts for the elevated local overdensity in the vicinity of the cluster host:
\begin{align}
\frac{dN_\mathrm{CDM}}{dM_h\,dV} = \delta_\mathrm{LOS}\,\xi(M_\mathrm{host},z)\,\left.\frac{dN}{dM_h\,dV}\right|_\mathrm{Sheth\text{-}Tormen},
\label{eq:los_mf}
\end{align}
where $\delta_\mathrm{LOS}$ is a normalization factor for the projected LOS volume element, and $\xi(M_\mathrm{host}, z)$ is the two-halo term, which captures the enhancement of the line-of-sight halo abundance from large-scale structure correlated with the massive cluster host (\citealt{Gilman2020}). We integrate over a double-cone geometry centered on the cluster.
Subhalos within the cluster are described by the projected subhalo mass function (SHMF), which follows a power-law in infall mass,
\begin{align}
\frac{dN_\mathrm{CDM}}{dM_h\,dA} = \frac{\Sigma_\mathrm{sub}}{M_0}\left(\frac{M_h}{M_0}\right)^{-\alpha}\mathcal{F}(M_\mathrm{host},z),
\label{eq:sub_mf}
\end{align}
where $\Sigma_\mathrm{sub}$ is the projected subhalo surface mass density amplitude, $M_0$ is a pivot mass, $\alpha$ is the power-law slope, and $\mathcal{F}(M_\mathrm{host},z)$ encodes the dependence on the host mass and redshift.
We calibrate $\Sigma_\mathrm{sub}$ using the SHMF derived from strong lensing observations of Abell 2744 (\citealt{Natarajan2017}, Figure~4), and follow \citet{Gannon2025} for calibrations of the scaling with host mass and redshift.

Cluster subhalos undergo significant tidal stripping as they orbit within the cluster potential.
We model this with a probabilistic approach in which the bound mass fraction $f_\mathrm{bound} \equiv m_\mathrm{bound}/m_\mathrm{infall}$ is drawn independently for each subhalo from a log-normal distribution,
\begin{align}
\log_{10}f_\mathrm{bound} \sim \mathcal{N}\!\left(\mu=-1.0,\,\sigma=0.5\right),
\label{eq:fbound}
\end{align}
so that subhalos retain on average $\sim\!10\%$ of their infall mass following tidal disruption (\citealt{Du2025hbv}).
The density profile of each stripped subhalo is then modeled as a tidally truncated NFW halo, $\rho\left(r\right) = \rho_{\rm{NFW}}\left(r\right) f_t/(1+r^2 / r_t^2)$, where $r_t$ is the tidal truncation radius and $f_t$ a normalization factor, both fixed by the drawn bound mass fraction, following the prescription implemented in pyHalo that is based on simulations presented by \citealt{Du2024}.

\subsubsection{Wave Dark Matter ($\psi$DM).}

For $\psi$DM, we use pyHalo's \texttt{ULDM} class (\citealt{Laroche2022}), which superimposes density fluctuations on the CDM subhalo population described above, applying the same tidal-stripping prescription.
In $\psi$DM, the ultra-light boson of mass $m_\psi$ behaves as a coherent classical wave on astrophysical scales. Wave interference between wave modes produces stochastic density fluctuations that are spatially correlated on the de Broglie wavelength,
\begin{align}
\lambda_\mathrm{dB} = 0.6\left(\frac{m_\psi}{10^{-22}\,\mathrm{eV}}\right)^{-1}\!\left(\frac{v}{200\,\mathrm{km\,s^{-1}}}\right)^{-1}\,\mathrm{kpc},
\label{eq:debroglie}
\end{align}
where $v$ is the virial velocity of the host cluster halo at the lens redshift, computed from the cluster mass and the critical density at $z_\mathrm{lens}$.
Equation~\ref{eq:debroglie} shows that more massive bosons produce shorter-wavelength, finer-grained fluctuations in the convergence field.

Free-streaming of the $\psi$DM boson also suppresses structure formation on small scales.
The half-mode mass,
\begin{align}
M_{1/2} = 3.8\times10^{10}\left(\frac{m_\psi}{10^{-22}\,\mathrm{eV}}\right)^{-4/3}\!M_\odot,
\label{eq:half_mode}
\end{align}
characterizes the halo mass below which the $\psi$DM matter power spectrum is suppressed by half relative to CDM (\citealt{Schive2016,Hui2017}).
The redshift-dependent minimum halo mass is
\begin{align}
M_\mathrm{min}(z) = 1.2\times10^8\left(\frac{m_\psi}{10^{-22}\,\mathrm{eV}}\right)^{-3/2}(1+z)^{3/4}\left(\frac{\zeta(z)}{\zeta(0)}\right)^{1/4}\!M_\odot,
\label{eq:mmin}
\end{align}
where $\zeta(z)$ is a redshift-dependent function related to the linear growth of structure, and is set by the balance between the Jeans pressure and gravity \citep{Schive2016}.
This mass scale sets the lower integration limit of the halo mass function (Eq.~\ref{eq:los_mf}).

The amplitude of the density fluctuations is parametrized by $\log_{10}A_\psi$, a free parameter that sets the typical amplitude of a fluctuation in the projected mass associated with the wave interference effects of ULDM.
Following the notation of \citealt{Laroche2022}, $\log_{10}A_\psi = \log_{10}A_\mathrm{fluc}/2$, where $A_\mathrm{fluc}$ is the fluctuation amplitude that is calibrated against numerical simulations of wave interference effects in a ULDM halo presented by \citealt{Yavetz2022}.
Their physical prediction for galaxy-scale halos, accounting for the dark matter fraction at the Einstein radius, gives  $\log_{10}A_\psi = -0.8$. We adopt this as our reference physical amplitude throughout.
Each fluctuation is modeled as a Gaussian with zero mean, so that the rendered field perturbs the convergence around the smooth host density profile without altering its mean. The number of fluctuations per unit area scales as $\lambda_\mathrm{dB}^{-2}$, since each has a characteristic size set by the de Broglie wavelength.
Individual fluctuations are rendered stochastically within the lensing aperture, with the number of rendered fluctuations chosen to be large enough that the statistical properties of the resulting convergence field are converged.

\begin{figure*}
    \centering
    \includegraphics[width=\linewidth]{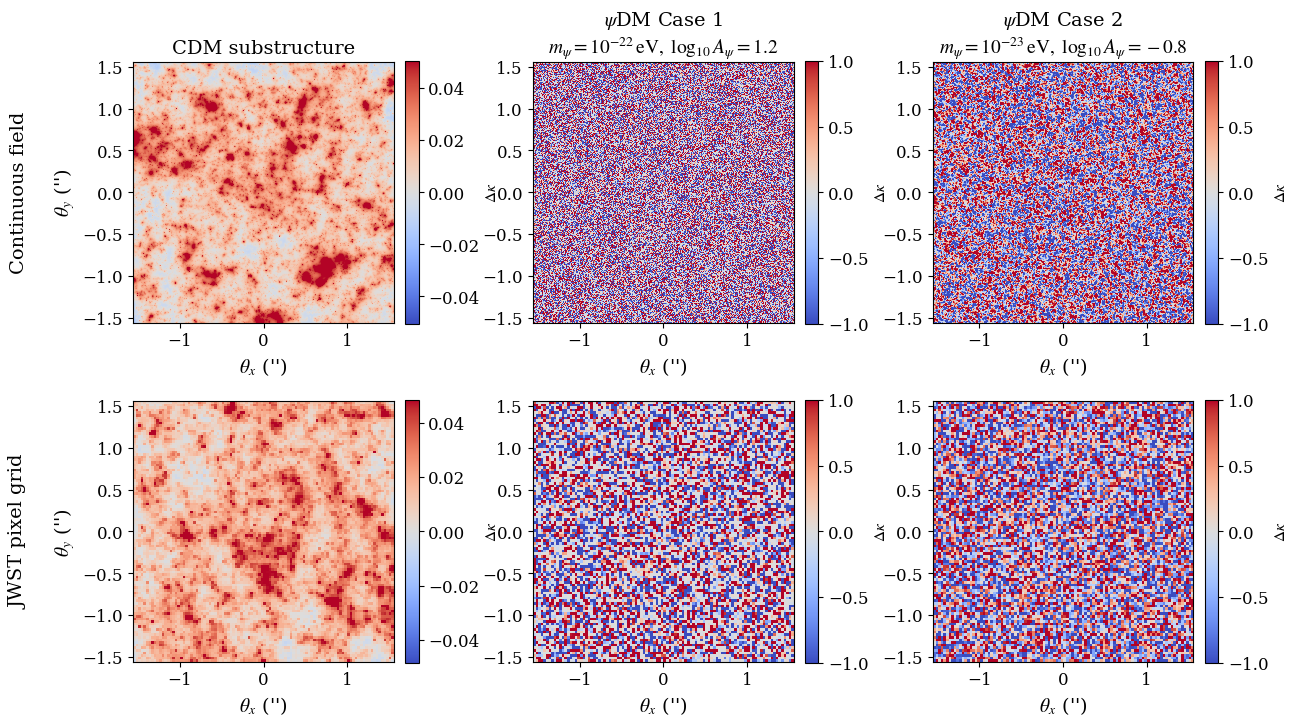}
    \caption{Projected substructure convergence fields $\Delta \kappa$ for a representative mock realization under three dark matter models. \textit{Columns} (left to right): CDM, showing perturbations from subhalos with masses between $10^{6}-10^{9}\,M_{\odot}$; $\psi$DM Case~1 ($m_\psi = 10^{-22}\,\mathrm{eV}$, $\log_{10}A_\psi = 1.2$), showing high-amplitude density fluctuations on the $\lambda_\mathrm{dB}$ scale; and $\psi$DM Case~2 ($m_\psi = 10^{-23}\,\mathrm{eV}$, $\log_{10}A_\psi = -0.8$), showing lower-amplitude fluctuations on a larger spatial scale, reflecting the longer $\lambda_\mathrm{dB}$ of the lighter boson. \textit{Top row}: the continuous physical $\Delta\kappa$ field, evaluated on a grid with a pixel scale $8\times$ finer than JWST. \textit{Bottom row}: the same field sampled on the JWST pixel grid ($100 \times 100$ pixels at $0.031''\,\mathrm{pixel}^{-1}$) used for the mock observations.}
    \label{fig:convergence_fields}
\end{figure*}

Figure~\ref{fig:convergence_fields} illustrates the projected convergence fields for CDM and the two $\psi$DM scenarios considered in this work (explained below).
The CDM field is dominated by the smooth NFW cluster profile and shows only mild, small-scale perturbations from the subhalo population.
In contrast, both $\psi$DM cases exhibit density fluctuations superimposed on the smooth background.
Case~1 ($m_\psi = 10^{-22}\,\mathrm{eV}$, $\log_{10}A_\psi = 1.2$, i.e.\ $100\times$ the physical prediction) produces compact, high-amplitude oscillations set by the shorter de Broglie wavelength of the heavier boson.
Case~2 ($m_\psi = 10^{-23}\,\mathrm{eV}$, $\log_{10}A_\psi = -0.8$, the physical prediction) displays lower-amplitude fluctuations on a larger spatial scale, consistent with the de Broglie wavelength growing as $\lambda_\mathrm{dB} \propto m_\psi^{-1}$ (Eq.~\ref{eq:debroglie}). At the cluster virial velocity, this corresponds to $\lambda_\mathrm{dB} \approx 0.78\,\mathrm{kpc}$ (${\sim}5$ pixels), compared with the sub-pixel $\lambda_\mathrm{dB} \approx 0.08\,\mathrm{kpc}$ of Case~1.

\subsection{JWST-quality simulations}
\label{sec:jwst_sims}

We simulate JWST-quality mock observations using \texttt{lenstronomy} (\citealt{Birrer2021_lenstronomy}) configured for the NIRCam F150W filter, which provides the finest available pixel scale, a sharp point spread function (PSF), and a strong sensitivity to $z \sim 1$ background galaxies.
The instrumental and observational parameters are summarized in Table~\ref{tab:jwst_obs_settings}.

\textbf{Pixel scale and image geometry:}
We adopt the NIRCam short-wavelength channel drizzled pixel scale of $0.0312''$\,pixel$^{-1}$.
Each mock image is $100\times100$ pixels, corresponding to a field of view of ${\approx}\,3.1''\times3.1''$ centered on the lensed arc system.
Ray-tracing is performed tile by tile using \texttt{lenstronomy}'s \texttt{SimAPI}, after which a single PSF convolution is applied to the full image for computational efficiency.

\textbf{Point spread function:}
The PSF is modeled using \texttt{WebbPSF}\footnote{\url{https://www.stsci.edu/jwst/science-planning/proposal-planning-toolbox/psf-simulation-tool}}, the official STScI PSF simulation tool for JWST instruments.
We simulate the NIRCam F150W PSF at native pixel resolution and adopt a full-width at half-maximum (FWHM) of $0.05''$ from the NIRCam PSF performance documentation.
The resulting PSF kernel is stored as a pixelized array and passed directly to \texttt{lenstronomy} for convolution.

\textbf{Noise model:}
The noise in each pixel combines three contributions: (i) Poisson photon noise from the lensed source flux, (ii) sky background from zodiacal and thermal emission, and (iii) detector read noise.
The sky background rate of $10.26\,\mathrm{e^-\,s^{-1}}$ per $362.49$ pixels is taken from the JWST Exposure Time Calculator (ETC)\footnote{\url{https://jwst.etc.stsci.edu/}}, and the AB magnitude zero-point of $28.0$ follows the NIRCam F150W flux calibration (\citealt{Rigby2023}).
The read noise of $0.94\,\mathrm{e^-}$ per read and detector gain of $2.05\,\mathrm{e^-\,ADU^{-1}}$ are taken from the NIRCam detector performance specifications.

\textbf{Exposure time:}
Our fiducial mocks assume a total exposure time of $t_\mathrm{exp} = 72{,}000\,\mathrm{s}$ ($20\,\mathrm{hours}$), representative of a deep JWST cluster lensing program.

\begin{table}
    \centering
    \begin{tabular}{lll}
        \hline
        Parameter & Value & Source \\
        \hline
        Filter & NIRCam F150W & --- \\
        Pixel scale & $0.0312''$\,pixel$^{-1}$ & drizzled SW channel \\
        Image size & $100\times100$ pixels & --- \\
        Field of view & ${\approx}3.1''\times3.1''$ & --- \\
        PSF model & \texttt{WebbPSF} (pixelised) & \citealt{Perrin2014} \\
        PSF FWHM & $0.050''$ & NIRCam PSF docs \\
        Exposure time & $72{,}000\,\mathrm{s}$ ($20\,\mathrm{hr}$) & --- \\
        Sky background & $10.26\,\mathrm{e^-\,s^{-1}}$ per $362.49$ pix & JWST ETC \\
        AB zero-point & $28.0\,\mathrm{mag}$ & \citealt{Rigby2023} \\
        Read noise & $0.94\,\mathrm{e^-\,read^{-1}}$ & NIRCam detector specs \\
        Detector gain & $2.05\,\mathrm{e^-\,ADU^{-1}}$ & NIRCam detector specs \\
        \hline
    \end{tabular}
     \caption{NIRCam F150W observational and detector settings used in the JWST mock simulations.}
      \label{tab:jwst_obs_settings}
\end{table}

\subsection{Measuring $P_{\delta}(k)$}
\label{sec:measuring_ps}

The residual power spectrum $P_\delta(k)$ is the two-dimensional power spectrum of the normalized residual images obtained after subtracting the best-fit smooth lens and source model from each observed (mock) image.
Dark matter substructure perturbs the lensing deflection field on scales set by the projected masses and positions of subhalos, and perturbations that cannot be absorbed by the smooth CAB lens model manifest as spatially coherent patterns in the residuals.
The amplitude and scale-dependence of $P_\delta(k)$ therefore encode the statistical properties of the underlying subhalo population.

The discriminating power against $\psi$DM arises from the coherence of the density fluctuations.
The de Broglie interference pattern produces density perturbations correlated on $\lambda_\mathrm{dB}$ (Eq.~\ref{eq:debroglie}), which imprint excess power in $P_\delta(k)$ at wavenumbers $k \lesssim 2\pi/\lambda_\mathrm{dB}$ relative to CDM.
In projection, contributions from a range of coherence lengths along the line of sight smear this excess into a broadband excess rather than a narrow spectral peak. The de Broglie scale therefore sets the characteristic wavenumber \emph{above} which the excess power falls off, rather than producing a peak at a single $k$.
Lighter bosons have larger $\lambda_\mathrm{dB}$ and consequently shift the excess to smaller $k$ (larger scales), while heavier bosons confine it to larger $k$, providing a handle on $m_\psi$.

We measure $P_\delta(k)$ from each mock observation through three sequential steps:
(i) model the source light with a shapelet basis (Section~\ref{sec:source_modeling}),
(ii) fit the local lens model jointly with the source via Bayesian nested sampling (Sections~\ref{sec:lens_modeling}--\ref{sec:nested_sampling}),
(iii) compute the 2D power spectrum of the normalized residual maps (Section~\ref{sec:power_spectra}).
Averaging $P_\delta(k)$ over many source realizations and substructure draws suppresses variance from individual source morphologies and stochastic halo realizations, making the population-level imprint of the dark matter model visible.

\subsection{Source modeling}
\label{sec:source_modeling}

We model the source light distribution using a shapelet basis (\citealt{Refregier2003}, \citealt{Birrer2015}).
Shapelets are an orthonormal set of two-dimensional basis functions constructed from weighted Hermite polynomials and form a complete basis for representing arbitrary smooth surface brightness distributions.
A shapelet model is characterized by a maximum polynomial order $n_\mathrm{max}$, a reference scale $\beta$ that sets the physical size of the basis, and a centroid $(\theta_{s,x}, \theta_{s,y})$.
The basis contains $m$ independent functions, whose coefficients are determined analytically by a linear solve at each likelihood evaluation (see Section~\ref{sec:nested_sampling}), where $m$ is set by:
\begin{align}
    m = \frac{(n_\mathrm{max}+1)(n_\mathrm{max}+2)}{2}.
    \label{eq:shapelet_nbasis}
\end{align}
The smallest spatial scale resolved at order $n_\mathrm{max}$ and scale $\beta$ is $\ell_\mathrm{min} = \beta/\sqrt{n_\mathrm{max}+1}$.
We reparametrize the shapelet scale as $\tilde{\beta} = \beta \sqrt{n_\mathrm{max}+1}$, where $\tilde{\beta}$ is sampled as a nonlinear parameter, ensuring $\ell_\mathrm{min}$ remains approximately constant with $n_\mathrm{max}$.
The source centroid $(\theta_{s,x}, \theta_{s,y})$ is also treated as a nonlinear free parameter with a uniform prior of $\pm0.2''$ around the image center.
A single shapelet model is shared across all three lensed images in the joint fit.
We adopt $n_\mathrm{max} = 20$ for the shapelet source, ensuring that the model is flexible enough to capture realistic COSMOS sources, while being computationally efficient. At the same time, a model that is overly flexible may risk absorbing some lensing perturbations from substructure, although a single source model constrained from three distinct images has limited freedom to mimic spatially incoherent image-plane residuals. We discuss this limitation in Section~\ref{sec:discussion}.

\subsection{Lens modeling}
\label{sec:lens_modeling}

We model the gravitational lens at each image position using the Curved Arc Basis (CAB) formalism (\citealt{Birrer2021}), which provides a local approximation to the lensing deflection field in the neighborhood of each image.
Rather than fitting a global parametric mass model for the full cluster, the CAB formalism describes the lensing effect at a given position through four observables derived from the local lensing Jacobian: the tangential stretch $\lambda_\mathrm{tan}$, the radial stretch $\lambda_\mathrm{rad}$, the arc curvature $s_\mathrm{tan}$, and the arc orientation $\phi$.

We use the \texttt{CURVED\_ARC\_SIS\_MST} implementation in \texttt{lenstronomy} (\citealt{Birrer2021_lenstronomy}), in which the local deflection angle at position $\vec{\theta}$ relative to a reference point $\vec{\theta}_0$ is
\begin{align}
\label{eq:cab_alpha}
    \vec{\alpha}(\vec{\theta})=\lambda_{\mathrm{rad}}^{-1}\!\left[\vec{\alpha}_{\mathrm{SIS}}(\vec{\theta})-\vec{\alpha}_{\mathrm{SIS}}\!\left(\vec{\theta}_0\right)\right]+\left(1-\lambda_{\mathrm{rad}}^{-1}\right)\!\left(\vec{\theta}-\vec{\theta}_0\right),
\end{align}
where $\vec{\theta}_0$ maps to the arc center by construction, and the SIS deflection centered at $\vec{\theta}_c$ is
\begin{align}
\label{eq:sis_alpha}
    \vec{\alpha}_{\mathrm{SIS}}(\vec{\theta})=s_{\mathrm{tan}}^{-1}\!\left(1-\frac{\lambda_{\mathrm{rad}}}{\lambda_{\mathrm{tan}}}\right)\frac{\vec{\theta}-\vec{\theta}_c}{|\vec{\theta}-\vec{\theta}_c|}.
\end{align}
The four CAB parameters $(\lambda_\mathrm{tan},\, \lambda_\mathrm{rad},\, s_\mathrm{tan},\, \phi)$ are independent for each image and can be estimated from the smooth NFW macro-model (Section~\ref{sec:dm_substructure}) by computing the local lensing Jacobian analytically at each image position.
We additionally include a two-component positional shift $(\alpha_x, \alpha_y)$ for each image to absorb rigid astrometric offsets not captured by the local arc model.

We perform a joint fit to all three lensed images simultaneously.
For image 1, we fix $\lambda_\mathrm{rad} = 1$ (normalizing the overall magnification scale), leaving three free CAB parameters. Images 2 and 3 each have four free CAB parameters plus two shift components.
Combined with the three nonlinear source parameters ($\tilde{\beta}$, $\theta_{s,x}$, $\theta_{s,y}$), the total number of nonlinear parameters is 18.

\subsection{Bayesian nested sampling}
\label{sec:nested_sampling}

We infer the joint posterior over all nonlinear model parameters $\mathbf{q}$ using Bayesian nested sampling implemented in \texttt{dynesty} (\citealt{Speagle2020}).
We adopt uniform priors on all nonlinear CAB parameters, centered on the smooth-model estimates with widths much larger than the lens parameter shifts expected from the addition of substructure:
\begin{align*}
    \lambda_\mathrm{rad} &\in \hat{\lambda}_\mathrm{rad} \pm 1.5, &
    \lambda_\mathrm{tan} &\in \hat{\lambda}_\mathrm{tan} \pm 2.25, \\
    s_\mathrm{tan} &\in \hat{s}_\mathrm{tan} \pm 0.75, &
    \phi &\in \hat{\phi} \pm \tfrac{3\pi}{16},
\end{align*}
where hatted quantities denote the NFW macro-model estimates. Positional shifts are constrained to $|\alpha_{x,y}| \leq 0.2''$, and the source scale parameter to $\tilde{\beta} \in [0.0,\, 2.5]$ arcsec.

At each likelihood evaluation, the $m = 231$ linear shapelet coefficients are solved for analytically via weighted least squares (WLS) by constructing the joint linear response matrix across all three images and inverting it against the pixel data (\citealt{Birrer2015}).
This semi-linear approach reduces the effective nonlinear parameter space from $18 + 231$ to 18 dimensions, dramatically accelerating convergence.
The pixel-level Gaussian likelihood is
\begin{align}
    P(\mathbf{D} \mid \mathbf{q})=\frac{\exp \!\left[-\frac{1}{2}(\mathbf{D}-\mathbf{M}(\mathbf{q}))^T \boldsymbol{\Sigma}_{\mathrm{pixel}}^{-1}(\mathbf{D}-\mathbf{M}(\mathbf{q}))\right]}{\sqrt{(2 \pi)^{\dim(\mathbf{D})} \det\!\left(\boldsymbol{\Sigma}_{\mathrm{pixel}}\right)}},
    \label{eq:likelihood}
\end{align}
where $\mathbf{D}$ is the vector of observed pixel fluxes across all three images, $\mathbf{M}(\mathbf{q})$ is the model prediction, and $\boldsymbol{\Sigma}_\mathrm{pixel}$ is the diagonal noise covariance matrix with per-pixel variances combining Poisson photon noise and read noise in quadrature.
The posterior is computed using Bayes' theorem:
\begin{align}
    P(\mathbf{q} \mid \mathbf{D}) \propto P(\mathbf{q})\, P(\mathbf{D} \mid \mathbf{q}).
    \label{eq:posterior}
\end{align}

We run the nested sampler with a convergence criterion of $\Delta\ln\mathcal{Z} \leq 10^{-3}$, where $\mathcal{Z} = \int P(\mathbf{q})\,P(\mathbf{D}\mid\mathbf{q})\,d\mathbf{q}$ is the Bayesian evidence and $\Delta\ln\mathcal{Z}$ is the estimated remaining change in $\ln\mathcal{Z}$ from continued iteration.
Sampling is parallelized across 112 CPU cores, with each core evaluating an independent likelihood call.
For each dark matter model and source realization, we save the full posterior sample, from which we subsequently compute the best-fit image reconstruction and residual map.

\subsection{Power spectrum computation}
\label{sec:power_spectra}

From each of the $N_\mathrm{src} = 20$ source realizations per dark matter model, we obtain three residual images -- one per lensed image -- giving $N_\mathrm{maps} = 60$ residual maps per model.
Each residual map is defined as the noise-normalized difference between the observed image and the best-fit model at the posterior maximum $\hat{\mathbf{q}}$:
\begin{align}
    \delta_i(\vec{\theta}) = \frac{D_i(\vec{\theta}) - M_i\!\left(\hat{\mathbf{q}},\,\vec{\theta}\right)}{\sigma_i(\vec{\theta})},
    \label{eq:residual}
\end{align}
where $D_i$ is the observed pixel flux, $M_i(\hat{\mathbf{q}})$ is the model-predicted flux, and $\sigma_i$ is the per-pixel noise standard deviation combining Poisson and read noise in quadrature.
Normalizing by $\sigma_i$ places residuals in units of noise standard deviations, so that the power spectrum of pure noise is a flat (white) spectrum. The excess power above this white-noise floor signals coherent structure in the residuals from any combination of unmodeled substructure and imperfections in the source\,+\,lens modeling. We isolate the latter in Appendix~\ref{app:systematics}. In what follows, we refer to the excess from systematics as the \emph{modeling residual floor} and to genuine substructure-driven excess above this floor as the substructure signal.

We compute $P_\delta(k)$ for each residual map using the \texttt{powerbox} Python package (\citealt{Murray2018}).
For each $100\times100$ pixel map, \texttt{powerbox} performs a 2D discrete Fourier transform and azimuthally averages $|\tilde{\delta}(k_x, k_y)|^2$ into 20 linearly-spaced annular bins, returning the isotropic power spectrum $P_\delta(k)$.
To convert the Fourier wavenumber from pixel$^{-1}$ to physical units, we project the image-plane pixel scale to the lens plane.
The physical field size is
\begin{align}
    L = N_\mathrm{pix}\,\theta_\mathrm{pix}\,d_A^\mathrm{lens},
\end{align}
where $N_\mathrm{pix} = 100$, $\theta_\mathrm{pix} = 0.031''$\,pixel$^{-1}$, and $d_A^\mathrm{lens}$ is the angular diameter distance to the lens at $z_\mathrm{lens} = 0.4$, computed using \texttt{lenstronomy}'s \texttt{LensCosmo} module with Planck cosmology (\citealt{Planck2020}).
The resulting $k$-bins span $k \in [0.85,\, 25]\,\mathrm{kpc}^{-1}$, corresponding to spatial scales of approximately $0.25$--$7\,\mathrm{kpc}$ in the lens plane.

For each dark matter model, we report the median $P_\delta(k)$ across all 60 residual maps, with the 16th–84th percentile band as a measure of variance due to source morphology and subhalo stochasticity.
The ensemble comparison of $P_\delta(k)$ between CDM and $\psi$DM constitutes our primary observable for dark matter model discrimination.

\section{Results}
\label{sec:results}

We present results for two complementary analyses.
First, as a theoretical upper bound, we estimate the residual power spectrum $P_\delta(k)$ directly from the mock images without any lens modeling, to establish what signal is in principle available.
Second, we apply the full CAB\,+\,shapelet fitting pipeline to 20-hour JWST mock observations, computing $P_\delta(k)$ from the fitting residuals for three dark matter models: CDM, and two $\psi$DM scenarios that bracket the range of observational interest.
\textbf{Case 1} uses $m_\psi = 10^{-22}\,\mathrm{eV}$ with $\log_{10}A_\psi = 1.2$, i.e.\ $100\times$ the physical prediction of \citealt{Laroche2022}, probing a strongly detectable $\psi$DM signal.
\textbf{Case 2} uses $m_\psi = 10^{-23}\,\mathrm{eV}$ with $\log_{10}A_\psi = -0.8$, the physical prediction of \citealt{Laroche2022}, representing the most astrophysically motivated scenario.
To map the mass dependence at the physical amplitude, we additionally present results for $\log_{10}(m_\psi/\mathrm{eV}) \in \{-22, -21\}$ alongside Case~2 in the power spectrum comparison.
For each model we analyze $N_\mathrm{src} = 20$ source realizations $\times$ 3 images $= 60$ residual maps.

\subsection{Theoretical Residual Power Spectra}
\label{sec:results_theoretical}

As a theoretical baseline, we compute the residual power spectrum $P_\delta(k)$ directly from the raw lensing signal of DM substructure, without any modeling step.
For each mock realization, we generate a smooth image (no substructure) and a substructure image under the same source and macro-lens, and form the deviation map
\begin{equation}
    \delta_i^\mathrm{th} = \frac{I_i^\mathrm{subs} - I_i^\mathrm{smooth}}{\sigma_i},
\end{equation}
where $I_i^\mathrm{subs}$ and $I_i^\mathrm{smooth}$ are the substructure and smooth mock images respectively, and $\sigma_i$ is the pixel noise.
This quantity isolates the pure lensing imprint of DM substructure and, under the assumption of perfect knowledge of the smooth lens model, sets an upper bound on the discriminating power achievable by any fitting-based analysis.
While we add negative convergence mass sheets to compensate for the extra projected mass from substructure, the CDM subhalo population is clumpy and non-Gaussian, and can produce coherent large-scale perturbations to the effective smooth lens model. This introduces some inaccuracy into the theoretical residuals, which is reflected in the systematic offset between the best-fit and CAB-estimate parameters shown in Appendix~\ref{app:posteriors}.
In practice, this means the theoretical $P_\delta^\mathrm{th}(k)$ slightly overestimates the large-scale (low-$k$) power relative to the full fitting analysis: a realistic CAB fit absorbs part of this coherent large-scale perturbation into the smooth macro-model, whereas the theoretical residual, computed against the true input macro-model, retains it. The theoretical spectra should therefore be read as an upper bound on the available signal rather than an exact prediction of the fitted result.

\begin{figure}
    \centering
    \includegraphics[width=\linewidth]{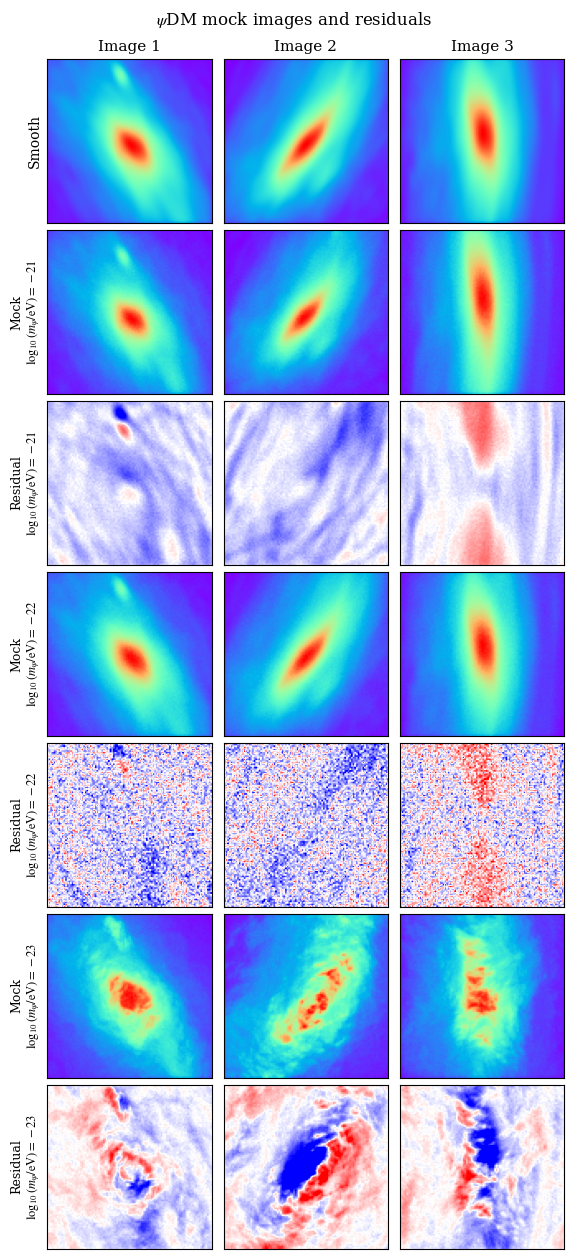}
    \caption{Deviation maps $\delta_i^\mathrm{th} = (I^\mathrm{subs} - I^\mathrm{smooth})/\sigma$ for $\psi$DM at three boson masses $\log_{10}(m_\psi/\mathrm{eV}) \in \{-21,\,-22,\,-23\}$ at the physical fluctuation amplitude $\log_{10}(A_\psi) = -0.8$ from \citealt{Laroche2022}, for a representative source realization. \textit{Top row}: smooth reference images. \textit{Subsequent row pairs}: each pair shows, for one boson mass, the $\psi$DM mock image (top of the pair) and deviation map (bottom of the pair, with $\pm 30\sigma$, $\pm 3\sigma$ and $\pm 30\sigma$ color scales, respectively). At $m_\psi = 10^{-21}\,\mathrm{eV}$, $\lambda_\mathrm{dB} \approx 0.008\,\mathrm{kpc}$ is sub-pixel at cluster virial velocities. Fluctuations are unresolved, and residuals are dominated by the CDM-like subhalo population. At $m_\psi = 10^{-22}\,\mathrm{eV}$, subhalos are strongly suppressed, but $\lambda_\mathrm{dB} \approx 0.08\,\mathrm{kpc}$ remains sub-pixel and the fluctuation amplitude is small, leaving only faint structure. At $m_\psi = 10^{-23}\,\mathrm{eV}$, $\lambda_\mathrm{dB} \approx 0.78\,\mathrm{kpc}$ (${\sim}5$ pixels) is resolved, and the granules produce the coherent, large-scale alternating residuals visible in the bottom row pair.}
    \label{fig:mass_reconstructions}
\end{figure}

\begin{figure}
    \centering
    \includegraphics[width=\linewidth]{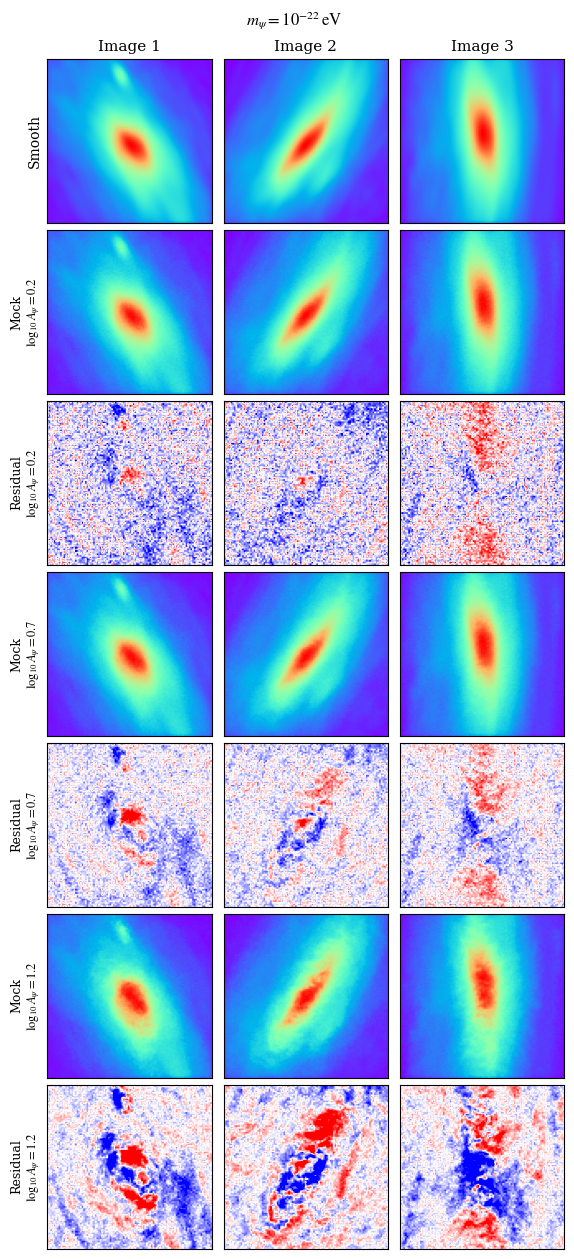}
    \caption{Deviation maps $\delta_i^\mathrm{th} = (I^\mathrm{subs} - I^\mathrm{smooth})/\sigma$ for $\psi$DM with fixed $m_\psi = 10^{-22}\,\mathrm{eV}$ at three fluctuation amplitudes $\log_{10}(A_\psi) \in \{0.2,\,0.7,\,1.2\}$, for a representative source realization. \textit{Top row}: smooth reference images. \textit{Subsequent row pairs}: each pair shows, for one fluctuation amplitude, the $\psi$DM mock image (top of the pair) and its deviation map (bottom of row pair) with $\pm 3\sigma$, $\pm 5\sigma$, and $\pm 7\sigma$ color scales, respectively for increasing $A_\psi$. At $\log_{10}(A_\psi) = 0.2$, the deviation maps are barely distinguishable from noise, suppressing residual structure even relative to CDM prediction. At $\log_{10}(A_\psi) = 0.7$, stronger structure begins to emerge in the deviation maps. At $\log_{10}(A_\psi) = 1.2$, the density fluctuations produce pronounced, spatially extended patches of alternating positive and negative signal, demonstrating that the detectability of $\psi$DM with $P_\delta(k)$ grows strongly with the fluctuation amplitude.}
    \label{fig:amp_reconstructions}
\end{figure}

A key parameter governing the $\psi$DM signal is the fluctuation amplitude $A_\psi$.
\citealt{Laroche2022} provide a physical prediction $\log_{10}A_\psi = -0.8$, calibrated against the numerical $\psi$DM simulations of \citet{Yavetz2022}. However, this calibration was derived for galaxy-scale halos, and its applicability to the dense environments of galaxy clusters is uncertain. In clusters, the coherence of $\psi$DM on kpc scales may be altered by tidal forces, baryonic feedback, and the substantially deeper gravitational potential, all of which could suppress or enhance the amplitude of density fluctuations relative to the field prediction.
We therefore treat $A_\psi$ as a free parameter in this work, and use the theoretical power spectra to understand how the detectability of $\psi$DM with $P_\delta(k)$ depends on both $A_\psi$ and the boson mass $m_\psi$.

Figure~\ref{fig:mass_reconstructions} shows the deviation maps $\delta_i^\mathrm{th}$ for three boson masses at the physical fluctuation amplitude $\log_{10}(A_\psi) = -0.8$ from \citealt{Laroche2022}, illustrating three distinct physical regimes set by the interplay between the subhalo abundance vs. wave-interference scale and amplitude in the cluster environment.

At $m_\psi = 10^{-21}\,\mathrm{eV}$, the de Broglie wavelength at the cluster virial velocity ($v_{200} \approx 1535\,\mathrm{km\,s^{-1}}$ for $M_{200} = 10^{15}\,M_\odot$ at $z=0.4$) is $\lambda_\mathrm{dB} \approx 0.008\,\mathrm{kpc}$ (Equation~\ref{eq:debroglie}), far below the pixel scale of ${\approx}0.17\,\mathrm{kpc}$.
The density fluctuations are therefore unresolved, contributing negligibly to the deviation maps.
At this mass, the halo mass function suppression is moderate, so the subhalo population is similar to that of CDM. Consequently, the deviation maps are dominated by CDM-like subhalo contributions.

At $m_\psi = 10^{-22}\,\mathrm{eV}$, the subhalo mass function begins to be suppressed strongly, but the fluctuation scale grows only to $\lambda_\mathrm{dB} \approx 0.08\,\mathrm{kpc}$ -- still sub-pixel at cluster virial velocities -- and the physical fluctuation amplitude at this mass is small.
The combined effect is a deviation map with only faint, barely discernible structure.

At $m_\psi = 10^{-23}\,\mathrm{eV}$, the inverse-mass scaling $\lambda_\mathrm{dB} \propto m_\psi^{-1}$ brings the fluctuation scale to $\lambda_\mathrm{dB} \approx 0.78\,\mathrm{kpc}$ (${\sim}5$ pixels) and the fluctuation amplitude is significantly stronger. The density granules are now spatially resolved, and their imprint on the deviation maps is dramatic: coherent patches of alternating positive and negative signal, spanning many pixels, trace the underlying $\psi$DM granule pattern across the arc.
This progression directly reflects the scaling of the de Broglie wavelength $\lambda_\mathrm{dB} \propto m_\psi^{-1}$: lighter bosons produce larger-scale coherent density fluctuations that leave stronger imprints on the lensed surface brightness.

Figure~\ref{fig:amp_reconstructions} shows the complementary dependence on fluctuation amplitude for a fixed boson mass $m_\psi = 10^{-22}\,\mathrm{eV}$.
At $\log_{10}(A_\psi) = 0.2$, the deviation maps look similar to noise and the mock images are indistinguishable from the smooth reference by eye.
Increasing the amplitude to $\log_{10}(A_\psi) = 0.7$ produces faint but coherent speckled structure in the deviation maps, while at $\log_{10}(A_\psi) = 1.2$ the residuals show pronounced large-scale alternating positive and negative patches, and the mock images themselves develop visible granular texture.
This illustrates that even though individual $\psi$DM density fluctuations are hidden in the noise, their statistical imprint on $P_\delta(k)$ grows monotonically with $A_\psi$.

\begin{figure}
    \centering
    \includegraphics[width=\linewidth]{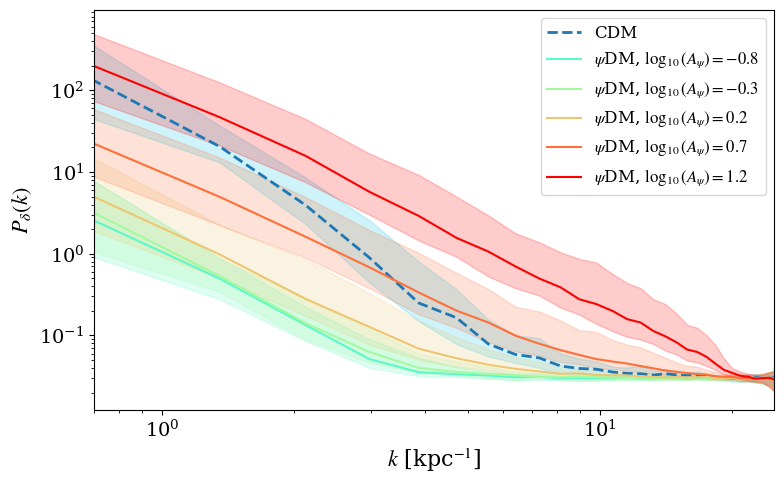}
    \caption{Theoretical residual power spectra $P_\delta^\mathrm{th}(k)$ for CDM and $\psi$DM with fixed boson mass $m_\psi = 10^{-22}\,\mathrm{eV}$ at five fluctuation amplitudes $\log_{10}(A_\psi) \in \{-0.8,\,-0.3,\,0.2,\,0.7,\,1.2\}$, computed from 60 deviation maps (20 source realizations $\times$ 3 images) per model. Solid lines show the median; shaded bands the 16th--84th percentile interval. All spectra decline steeply with $k$, and CDM and $\psi$DM differ in their slope of $P_{\delta}(k)$ at scales $k \lesssim 3\,\mathrm{kpc}^{-1}$. With $\log_{10}(A_\psi) = 1.2$, $\psi$DM produces a broadband elevation above the CDM baseline that is clearly distinguishable for $ k \gtrsim 3\,\mathrm{kpc}^{-1}$. Lower fluctuation amplitudes lead to a drop in power below the prediction of CDM.}
    \label{fig:ps_theoretical_amp}
\end{figure}

Figure~\ref{fig:ps_theoretical_amp} quantifies the amplitude dependence.
We show $P_\delta^\mathrm{th}(k)$ for CDM and $\psi$DM at five values of $\log_{10}(A_\psi) \in \{-0.8,\,-0.3,\,0.2,\,0.7,\,1.2\}$, spanning from $\sim\!1\times$ to $100\times$ the physical prediction ($\log_{10}A_\psi = -0.8$) of \citealt{Laroche2022}, computed from 60 deviation maps per model.
All spectra show a steeply declining power law, reflecting the combined effect of stochastic substructure perturbations, source morphology, local lens model, and the PSF.
At $\log_{10}(A_\psi) \in \{-0.8,\,-0.3,\,0.2\}$, the density fluctuations are weak, and because the CDM subhalo mass function is strongly suppressed at $m_\psi = 10^{-22}\,\mathrm{eV}$, the $\psi$DM signal falls \textit{below} the CDM prediction across the full $k$ range.
At $\log_{10}(A_\psi) = 0.7$, the growing fluctuations produce a $P_\delta(k)$ signal comparable in amplitude to CDM, though the two spectra differ in slope, reflecting the distinct spatial structure of $\psi$DM granules relative to discrete subhalos.
At $\log_{10}(A_\psi) = 1.2$, the $\psi$DM power dominates strongly over CDM, particularly at small scales, exceeding it by almost an order of magnitude at $k \sim 10\,\mathrm{kpc}^{-1}$.
Since the boson mass is fixed, the shape of the excess (above the white noise floor) does not change with amplitude -- only the overall normalization rises with increasing $A_{\psi}$.

\begin{figure}
    \centering
    \includegraphics[width=\linewidth]{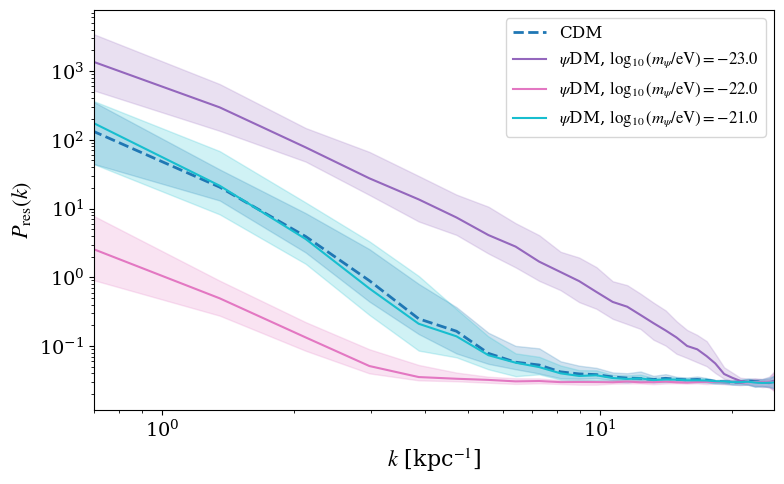}
    \caption{Theoretical residual power spectra $P_\delta^\mathrm{th}(k)$ for CDM and $\psi$DM at three boson masses $\log_{10}(m_\psi/\mathrm{eV}) \in \{-23, -22, -21\}$, with the fluctuation amplitude set to the physical prediction $\log_{10}(A_\psi) = -0.8$ from \citealt{Laroche2022}. Solid lines show the median over 60 deviation maps per model; shaded bands show the 16th--84th percentile interval. At the physical amplitude, $m_\psi = 10^{-23}\,\mathrm{eV}$ shows a clear excess above CDM at all scales, $m_\psi = 10^{-22}\,\mathrm{eV}$ shows less power than CDM at $k \lesssim 10$ kpc$^{-1}$, and $m_\psi = 10^{-21}\,\mathrm{eV}$ is indistinguishable from CDM.}
    \label{fig:ps_theoretical_all}
\end{figure}

In Figure~\ref{fig:ps_theoretical_all}, we show the dependence of $P_\delta^\mathrm{th}(k)$ on boson mass at the physical fluctuation amplitude $\log_{10}(A_\psi) = -0.8$ from \citet{Laroche2022}, which we note is an extrapolation to cluster environments.
We vary the boson mass over three values $\log_{10}(m_\psi/\mathrm{eV}) \in \{-23, -22, -21\}$, computed from 60 deviation maps per model.
At $m_\psi = 10^{-21}\,\mathrm{eV}$, the $\psi$DM spectrum is indistinguishable from CDM: the subhalo mass function suppression is moderate, while the $\lambda_\mathrm{dB}$ is far below the pixel scale and the fluctuation amplitude is negligibly small, so the density fluctuations do not contribute to the signal. At $m_\psi = 10^{-22}\,\mathrm{eV}$, the subhalo mass function is strongly suppressed but the fluctuation scale and amplitude remain small, so the net $P_\delta(k)$ signal drops \textit{below} the CDM prediction. Distinguishing this model from CDM would be possible with highly accurate source and lens modeling methods that can fit JWST-quality images with very low modeling systematics.
The clearest separation occurs at $m_\psi = 10^{-23}\,\mathrm{eV}$, where the larger $\lambda_\mathrm{dB}$ and stronger fluctuation amplitude produce a $\psi$DM spectrum that clearly exceeds CDM at most scales, with non-overlapping uncertainty bands.

\subsection{Joint Modeling: Reconstructions and Residuals}
\label{sec:results_reconstructions}

\begin{figure}
    \centering
    \includegraphics[width=\linewidth]{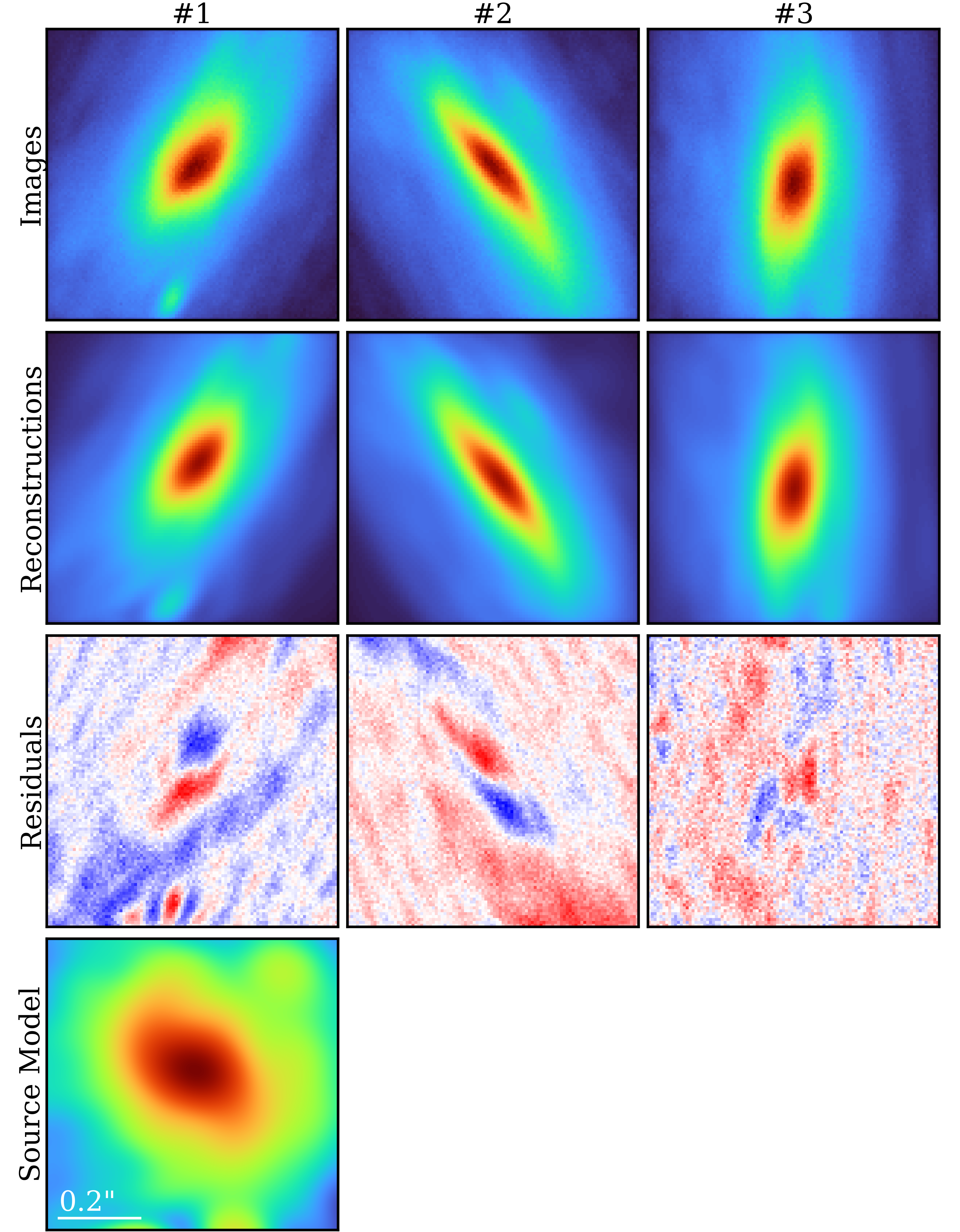}
    \caption{Representative 20-hour JWST mock observations of cluster-scale strong lensing with \textbf{CDM} substructure (top row), best-fit CAB\,+\,shapelet ($n_\mathrm{max}=20$) model reconstructions (second row), normalized residual maps $(D-M)/\sigma$ (third row), and reconstructed source model (bottom left), for each of the three lensed images. Residuals are low-amplitude (color map shows $\pm 3\sigma$). As we show in Appendix~\ref{app:systematics}, using a null test on smooth (no-substructure) mocks, the coherent residual structure visible here is dominated by source\,+\,lens modeling systematics rather than by lensing perturbations from CDM subhalos: at 20-hour JWST depth, with subhalos rendered in $10^6\le M_\mathrm{sub}/M_\odot \le 10^9$, the CDM signal sits at or below this modeling-systematics floor.}
    \label{fig:cdm_reconstructions}
\end{figure}

\begin{figure}
    \centering
    \includegraphics[width=\linewidth]{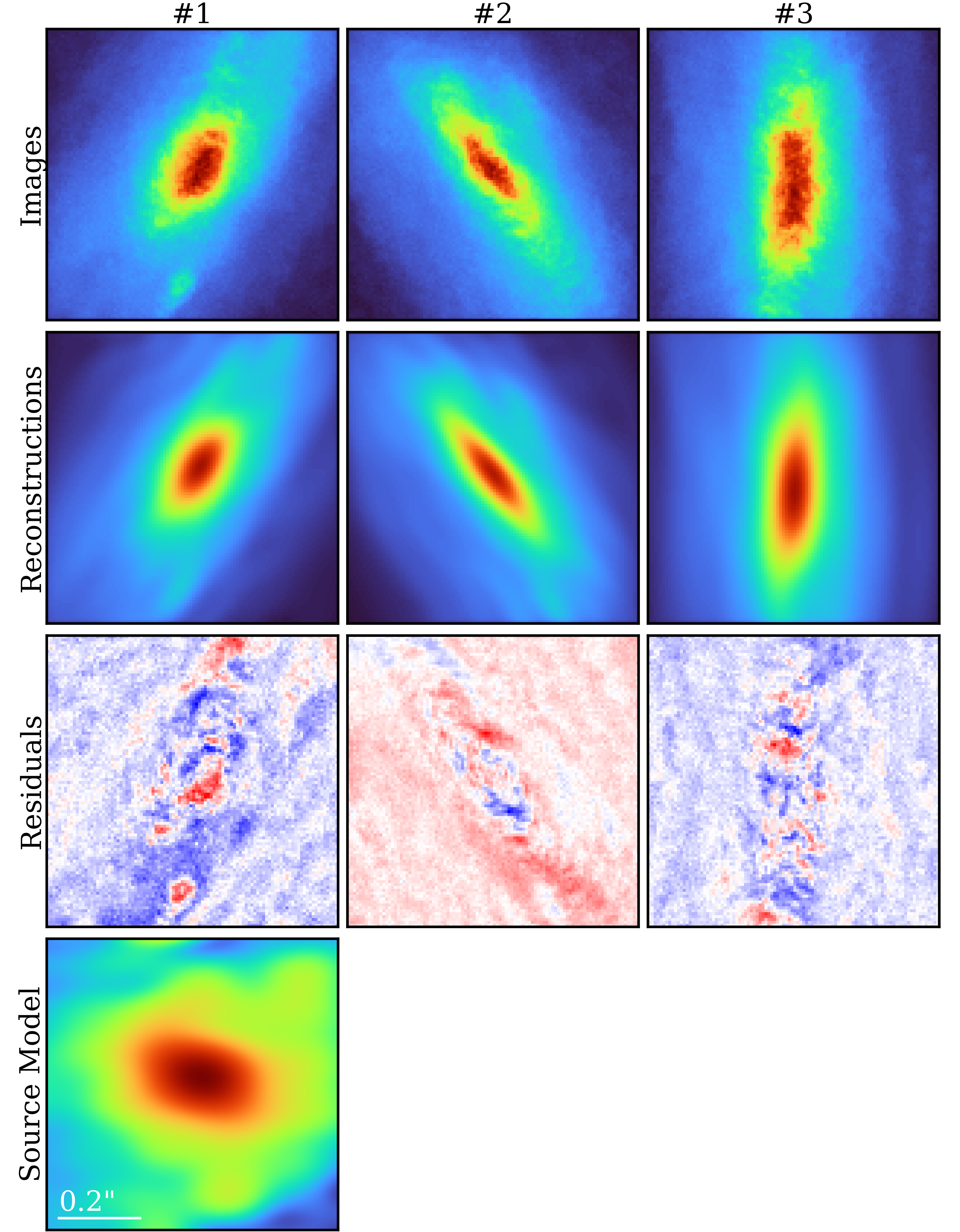}
    \caption{Same as Figure~\ref{fig:cdm_reconstructions}, but for \textbf{Case~1} $\psi$DM substructure: $m_\psi = 10^{-22}\,\mathrm{eV}$, $\log_{10}(A_\psi) = 1.2$ ($100\times$ the physical prediction of \citealt{Laroche2022}). The mock images show a subtle granular texture along the arcs compared to the smooth CDM case. The residual maps reveal coherent elongated patches of excess and deficit, preferentially aligned with the arc positions, at amplitudes of several $\sigma$.}
    \label{fig:uldm_case1_reconstructions}
\end{figure}

\begin{figure}
    \centering
    \includegraphics[width=\linewidth]{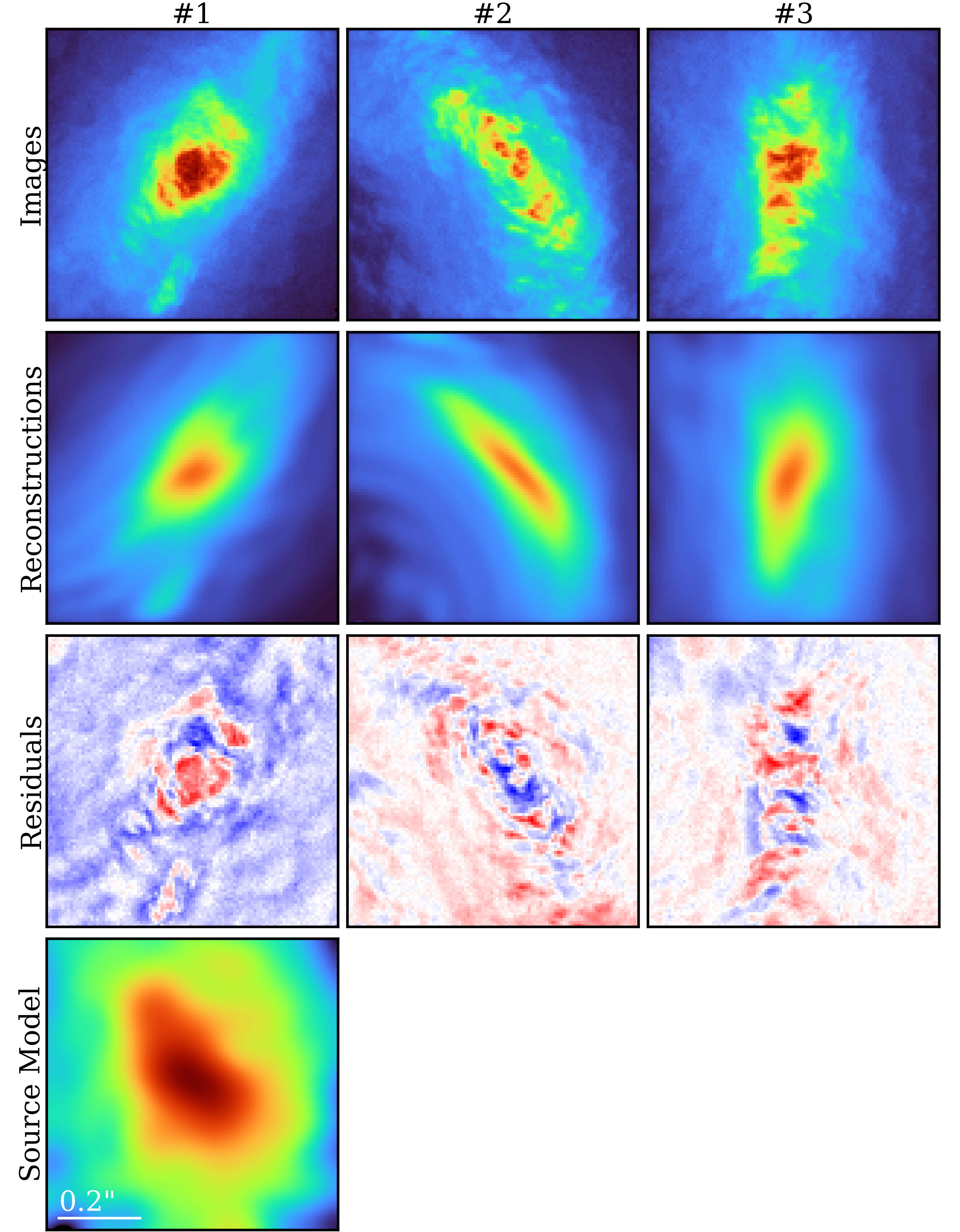}
    \caption{Same as Figure~\ref{fig:cdm_reconstructions}, but for \textbf{Case~2} $\psi$DM substructure: $m_\psi = 10^{-23}\,\mathrm{eV}$, $\log_{10}(A_\psi) = -0.8$ (the physical prediction of \citealt{Laroche2022}). The mock images show irregular, clumpy morphology compared to the smooth CDM case. The residual maps are dominated by large-scale coherent structures, with alternating patches of strong positive and negative residuals that trace the $\psi$DM density fluctuations on scales of several kpc, exceeding what is seen in the CDM case.}
    \label{fig:uldm_case2_reconstructions}
\end{figure}

Having established the theoretical baseline from direct image differencing, we now turn to the full fitting pipeline, in which the smooth lens model is not known but is inferred jointly with the source from each mock observation.
We first examine the visual appearance of the mock images and the quality of the CAB\,+\,shapelet fit for each dark matter model.
Figures~\ref{fig:cdm_reconstructions}--\ref{fig:uldm_case2_reconstructions} each show, for a representative source realization, the 20-hour JWST mock observation, the best-fit model reconstruction, and the normalized residual map $\delta_i = (D_i - M_i)/\sigma_i$ (Eq.~\ref{eq:residual}) for each of the three lensed images, together with the reconstructed source-plane model.

For CDM (Figure~\ref{fig:cdm_reconstructions}), the residual maps show low-amplitude coherent structure. A controlled null test on smooth, no-substructure mocks (Appendix~\ref{app:systematics}) reveals that this structure is dominated by source\,+\,lens modeling systematics rather than by lensing perturbations from CDM subhalos: the per-pixel signal from individual $10^6$--$10^9\,M_\odot$ subhalos sits below the floor set by imperfect source and lens modeling at our 20-hour JWST depth. The post-fit CDM residuals therefore measure the modeling-systematics floor -- the small-scale stochastic structure is the baseline against which the $\psi$DM signal will be assessed using the residual power spectrum.

For $\psi$DM Case~1 (Figure~\ref{fig:uldm_case1_reconstructions}), the residuals show a qualitative change in character relative to CDM.
The mock images exhibit a faint granular texture along the arc structures, which arise from the strong $\psi$DM density fluctuations at this high amplitude.
The reconstructions are smooth and do not capture these features, confirming that the $\psi$DM imprint is not absorbed into the macro-lens model but instead leaks into the residuals.

For $\psi$DM Case~2 (Figure~\ref{fig:uldm_case2_reconstructions}), at $m_\psi = 10^{-23}\,\mathrm{eV}$ and $\log_{10}(A_\psi) = -0.8$, the effect on the mock images is striking.
Unlike Case~1, the mock images themselves are visibly irregular: each arc shows a clumpy, non-smooth morphology that is visually distinguishable from the smooth CDM images.
The residual maps are dominated by large-scale coherent structures with alternating positive and negative regions, tracing the $\psi$DM density fluctuations on scales of several kpc. The source reconstruction also looks slightly more clumpy than Case 1, suggesting that some substructure signal may be absorbed by the flexible source model.
The marginalized posterior distributions of the CAB parameters recovered by the nested sampler for each dark matter model are presented in Appendix~\ref{app:posteriors}.

\subsection{Residual Power Spectra}
\label{sec:results_ps}

We now quantify the differences between the dark matter models using the residual power spectrum $P_\delta(k)$, computed from residuals obtained after strong-lensing analysis as described in Section~\ref{sec:power_spectra}.

\subsubsection{Case~1: $m_\psi = 10^{-22}\,\mathrm{eV}$, $100\times$ physical amplitude.}

\begin{figure}
    \centering
    \includegraphics[width=\linewidth]{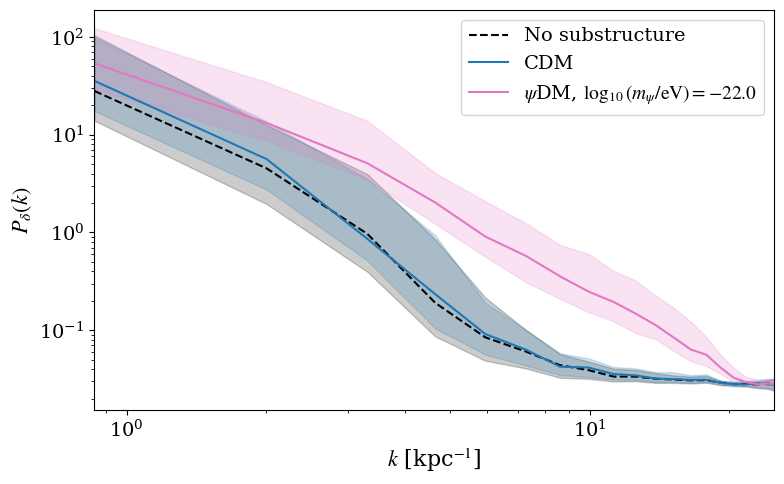}
    \caption{Residual power spectra $P_\delta(k)$ for CDM (blue), $\psi$DM Case~1 ($m_\psi = 10^{-22}\,\mathrm{eV}$ and $\log_{10}A_{\psi}=1.2$; pink), and smooth (see Appendix~\ref{app:systematics}; black) mocks. Colored lines show the median, and shaded bands show the 16th--84th percentile interval over 60 residual maps. The smooth and CDM curves coincide, showing that the CDM result is set by the modeling-systematics floor rather than by the lensing signal of subhalos.  The $\psi$DM median exceeds this baseline by a factor of $\sim\!3$--$10$ across $k \sim 2$--$11\,\mathrm{kpc}^{-1}$, with non-overlapping uncertainty bands over this range.}
    \label{fig:ps_case1}
\end{figure}

Figure~\ref{fig:ps_case1} shows $P_\delta(k)$ for CDM and $\psi$DM Case~1.
As shown in Appendix~\ref{app:systematics}, the CDM curve here is dominated by the modeling-systematics floor -- the residual power that the CAB\,+\,shapelet pipeline produces even on smooth, no-substructure mocks -- rather than by the bare lensing perturbations of individual CDM subhalos. The $\psi$DM Case~1 curve sits a factor of $\sim\!3$--$10$ above this floor across $k \sim 2$--$11\,\mathrm{kpc}^{-1}$, with non-overlapping 16th--84th percentile bands over this range, demonstrating clear discriminability of a strong $\psi$DM fluctuation amplitude from the CDM\,+\,systematics baseline in a single 20-hour JWST observation.
The shape of the $\psi$DM Case~1 spectrum -- a steep, featureless power law -- closely matches the corresponding theoretical prediction in Figure~\ref{fig:ps_theoretical_amp}, with a modest reduction in amplitude as the flexible CAB model absorbs some of the large-scale power. The CDM curve, by contrast, does not trace its theoretical counterpart: the systematics floor exceeds the bare CDM prediction at all scales shown, so the post-fit measurement is insensitive to the underlying subhalo signal in this mass range at our exposure depth.

\subsubsection{Case~2: physical fluctuation amplitude ($\log_{10}A_\psi = -0.8$), varying boson mass.}

\begin{figure}
    \centering
    \includegraphics[width=\linewidth]{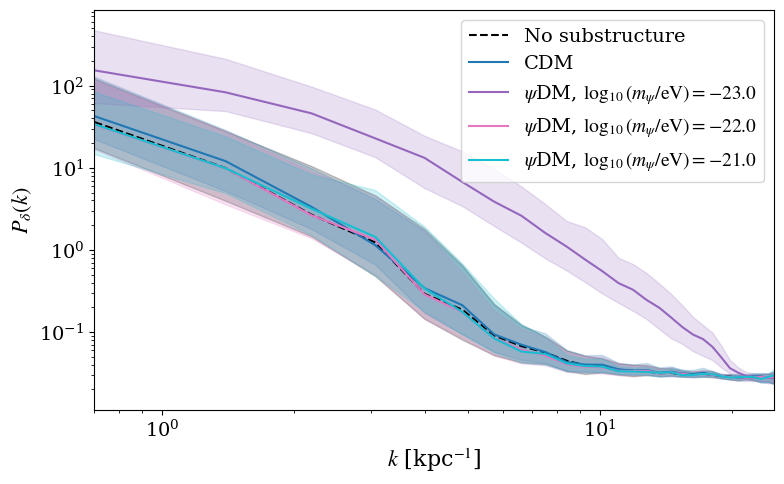}
    \caption{Residual power spectra $P_\delta(k)$ for CDM (blue), $\psi$DM at three boson masses $\log_{10}(m_\psi/\mathrm{eV}) \in \{-21, -22, -23\}$ with the fluctuation amplitude set to the physical prediction $\log_{10}A_\psi = -0.8$ from \citealt{Laroche2022}, and smooth mocks (see Appendix~\ref{app:systematics}; black). Colored lines show the median over 60 residual maps, and shaded bands show the 16th--84th percentile interval. The smooth, CDM and higher $\psi$DM mass ($m_\psi \gtrsim 10^{-22}\,\mathrm{eV}$) curves coincide, showing that the results are set by the modeling-systematics floor rather than by the lensing signal of substructure. The $m_\psi = 10^{-23}\,\mathrm{eV}$ (purple) is clearly separated from this baseline at all scales shown ($1\, \mathrm{kpc}^{-1} \lesssim k \lesssim 11\, \mathrm{kpc}^{-1}$).}
    \label{fig:ps_case2}
\end{figure}

Figure~\ref{fig:ps_case2} shows $P_\delta(k)$ for CDM and $\psi$DM at the physical fluctuation amplitude $\log_{10}(A_\psi) = -0.8$ from \citealt{Laroche2022} for three boson masses.
The overall shapes mirror those of the corresponding theoretical spectra in Figure~\ref{fig:ps_theoretical_all}, confirming that the fitting pipeline preserves the qualitative structure of the signal.
The $m_\psi = 10^{-23}\,\mathrm{eV}$ result shows a strong signal: the $\psi$DM band is clearly separated from CDM across all scales shown ($1\,\mathrm{kpc}^{-1} \lesssim k \lesssim 10\,\mathrm{kpc}^{-1}$), with non-overlapping uncertainty bands, demonstrating that the lightest boson mass probed is detectable with a single 20-hour JWST observation even at the physical fluctuation amplitude.
In contrast to the theoretical spectra (Figure~\ref{fig:ps_theoretical_all}), where $m_\psi = 10^{-22}\,\mathrm{eV}$ shows a significant suppression in power relative to CDM, both $m_\psi = 10^{-22}\,\mathrm{eV}$ and $m_\psi = 10^{-21}\,\mathrm{eV}$ are indistinguishable from CDM in the full fitting results.
This degradation reflects the impact of realistic modeling systematics: the fitting pipeline cannot reduce residuals to the noise floor, and the additional non-Gaussian residuals introduced by imperfect source and lens modeling erase the modest separations that are theoretically available at these masses.

\section{Discussion}
\label{sec:discussion}

Our results establish that the residual power spectrum $P_\delta(k)$ is a viable statistical probe of $\psi$DM in cluster lenses.
We begin with a physical interpretation of the broadband $P_\delta(k)$ excess and its connection to the substructure convergence power spectrum $P_\kappa(k)$ -- the power spectrum of the projected dark matter density field, first proposed as a lensing observable by \citet{Hezaveh2014} and computed analytically for a range of dark matter models (e.g. \citealt{Rivero2018b}).
We then situate our approach within the landscape of existing $\psi$DM constraints -- including flux-ratio anomalies and VLBI lensing -- and highlight the complementary regime probed by the cluster-scale power spectrum.
We subsequently discuss the limitations and modeling assumptions of the present work, and conclude by outlining the most promising directions for future work.

\subsection{Physical interpretation}

$P_\delta(k)$ is the observational realization of the substructure convergence power spectrum $P_\kappa(k)$ -- the power spectrum of the projected dark matter density field -- which \citet{Hezaveh2014} first proposed as a lensing observable and \citet{Rivero2018a, Rivero2018b} computed analytically for galaxy-scale systems, showing sensitivity to CDM, warm dark matter, and self-interacting dark matter. The present work extends this framework to the cluster regime and to the $\psi$DM model.
The mapping from $P_\kappa(k)$ to $P_\delta(k)$, however, involves several non-trivial steps, with contributions from the source galaxies, local lensing fields, and observational settings.
A key contribution of this paper is therefore to establish the quantitative transfer from the theoretically well-studied $P_\kappa(k)$ to the directly observable $P_\delta(k)$ in the cluster lensing context.

The results presented in Section~\ref{sec:results} establish two key points: the $\psi$DM signal survives the CAB\,+\,shapelet fitting step without being absorbed by the smooth lens or the flexible source model, and the resulting $P_\delta(k)$ is sensitive to both the boson mass and the fluctuation amplitude, with statistically significant separation from CDM achievable with a single 20-hour JWST observation.

The absence of a sharp spectral peak in $P_\delta(k)$ at $k \sim 2\pi/\lambda_\mathrm{dB}$ is physically expected in the regime of our simulations.
For $m_\psi = 10^{-22}\,\mathrm{eV}$ at the cluster virial velocity $v_{200} \approx 1535\,\mathrm{km\,s^{-1}}$, Equation~\ref{eq:debroglie} gives $\lambda_\mathrm{dB} \approx 0.08\,\mathrm{kpc}$, corresponding to $k \sim 80\,\mathrm{kpc}^{-1}$ -- well above the $k_\mathrm{max} \approx 25\,\mathrm{kpc}^{-1}$ of our analysis.
For $m_\psi = 10^{-23}\,\mathrm{eV}$, $\lambda_\mathrm{dB} \approx 0.78\,\mathrm{kpc}$ ($k \sim 8\,\mathrm{kpc}^{-1}$), which falls within our analysis range and is consistent with the stronger separation seen for this mass.
The broadband character of the $\psi$DM excess below the de Broglie scale reflects the two-dimensional projection of three-dimensional density fluctuations: the projected power spectrum receives contributions from all modes with $k_\perp \leq k$, smearing the de Broglie peak into a broad excess at lower wavenumbers.

\subsection{Comparison to existing constraints}

Our approach is complementary to existing $\psi$DM constraints from strong lensing.
Flux-ratio anomalies in quadruply lensed quasars \citep{Laroche2022} constrain the halo mass function suppression and host-halo fluctuations in field environments at $z \lesssim 1$.
VLBI lensing \citep{Powell2023} excludes $m_\psi \leq 4.4\times10^{-21}\,\mathrm{eV}$ by resolving individual granule structures, but requires milli-arcsecond radio imaging.
In this work, we instead study the \emph{statistical power spectrum} of optical surface brightness residuals in cluster lenses -- a statistical observable linking dark matter theories to their observational signatures.

The cluster regime offers specific advantages. Firstly, cluster lenses are the strongest known gravitational lenses, highly magnifying small statistical signals from substructure. The large lensing cross-section yields multiple, highly magnified arcs per system, and multiple independent source realizations in the cluster field of view, each contribute residual maps to the ensemble -- substantially reducing the variance of $P_\delta(k)$ relative to any single quad-lens system. On the other hand, the higher virial velocities of clusters ($v \sim 10^3\,\mathrm{km\,s^{-1}}$ vs.\ $\sim\!10^2\,\mathrm{km\,s^{-1}}$ for galaxy-scale halos) reduce $\lambda_\mathrm{dB}$ by an order of magnitude (Eq.~\ref{eq:debroglie}), which shifts the signal toward smaller scales for $m_\psi \gtrsim 10^{-22}\,\mathrm{eV}$ but leaves it accessible on kpc scales for $m_\psi \lesssim 10^{-23}\,\mathrm{eV}$. It may be possible to exploit compound lens configurations (e.g., a strong galaxy lens in a cluster lens) to overcome this issue and probe higher particle masses of $\psi$DM (see Section \ref{sec:future}).

\subsection{Limitations}

Several modeling assumptions and simplifications in the present work merit discussion.

\textit{Fluctuation amplitude in clusters.}
The most significant uncertainty in our analysis is the extrapolation of the \citealt{Laroche2022} fluctuation amplitude to cluster environments.
Equation~19 of \citealt{Laroche2022} was calibrated on cosmological $\psi$DM simulations of field halos. In clusters, the deeper gravitational potential may significantly modify the amplitude of the density fluctuations.
This uncertainty motivates treating $A_\psi$ as a free parameter rather than fixing it to the physical prediction, as we have done throughout.
The detectability thresholds derived here therefore bound the range of scenarios: physical-amplitude results represent the most conservative (astrophysically motivated) case, while the $100\times$ amplitude case represents a scenario in which cluster environments enhance the $\psi$DM fluctuation signal.

\textit{Single cluster configuration.}
All results are derived for a single cluster configuration ($M_{200} = 10^{15}\,M_\odot$, $c_{200} = 4$, $z_\mathrm{lens} = 0.4$).
Clusters spanning a range of masses, concentrations, ellipticities, and redshifts would probe different de Broglie scales and subhalo populations. The sensitivity of $P_\delta(k)$ to cluster parameters is left to future work.

\textit{CDM subhalo mass range.}
We exclude subhalos with $M_\mathrm{sub} > 10^9\,M_\odot$, which are individually detectable perturbers amenable to direct detection methods.
Including these objects would raise the CDM power spectrum at the lowest $k$-bins, potentially reducing the contrast with $\psi$DM.
A careful treatment of the high-mass subhalo contribution will be required in a real data application.

\textit{Modeling-systematics floor.}
At 20-hour JWST depth and over the rendered subhalo mass range, the post-fit CDM residual power spectrum is dominated by source\,+\,lens modeling systematics rather than by the bare lensing perturbations of CDM subhalos (Appendix~\ref{app:systematics}). The ${\psi}$DM-vs-CDM separations reported in \S\ref{sec:results_ps} are therefore separations between the $\psi$DM signal and a CDM\,+\,systematics floor. This is the relevant baseline for distinguishing $\psi$DM from a no-detection scenario in real data, but it also means that improving the source and lens modeling would lower the floor and enable detection of weaker substructure signals, including the bare CDM signal and $\psi$DM (or alternative DM model) signals that predict suppression of $P_{\delta}(k)$ relative to CDM.

\subsection{Future directions}
\label{sec:future}

The results presented here open several concrete directions for future work: extensions that probe new physical regimes, improvements to the statistical and modeling methodology, and broader observational programs that would sharpen the constraints.

An interesting extension of this work could exploit compound lens configurations in which a cluster member galaxy provides secondary strong lensing for a highly magnified background arc.
The relevant $\psi$DM fluctuations in such systems arise from the \emph{galaxy} halo, whose lower virial velocity ($v_{200} \approx 200\,\mathrm{km\,s^{-1}}$) yields a de Broglie wavelength $\approx 8\times$ larger than in the cluster halo ($v_{200} \approx 1535\,\mathrm{km\,s^{-1}}$) for the same boson mass.
At $m_\psi = 10^{-22}\,\mathrm{eV}$, this gives $\lambda_\mathrm{dB} \approx 0.6\,\mathrm{kpc}$ ($\approx 3$--$4$ JWST pixels) in the galaxy halo versus the sub-pixel $0.08\,\mathrm{kpc}$ in the cluster, extending the accessible mass range to $m_\psi \sim 10^{-22}$--$10^{-21}\,\mathrm{eV}$.
Although the galaxy-halo fluctuation amplitude is smaller, the extreme cluster magnification ($\mu \gtrsim 40$) boosts the photon counts and, thus, the signal-to-noise for small surface-brightness perturbations.
Such systems already exist in JWST data: for example, the host galaxy of SNe Requiem and Encore in MACS~J0138.0$-$2155 is multiply imaged by the cluster with additional strong lensing from a cluster-member galaxy \citep{Pierel2024}. Applying the residual power spectrum framework to such compound configurations would yield complementary $\psi$DM constraints at heavier boson masses.

While the power spectrum captures only Gaussian information, future work should explore higher-order statistics sensitive to the non-Gaussian, discrete nature of $\psi$DM granules.
The two-dimensional wavelet scattering transform \citep{BrunaMallat2013} is well suited to this task: by building a hierarchy of modulus wavelet coefficients, it retains phase information that $P_\delta(k)$ discards, and could yield substantially tighter joint constraints on $m_\psi$ and $A_\psi$.

The accuracy of the source reconstruction is another important direction for improvement.
The shapelet basis imposes a fixed angular-scale hierarchy that may miss the complex, clumpy morphologies common in high-redshift JWST sources.
Adaptive methods, such as Delaunay triangulation-based source reconstructions, can represent arbitrary morphologies with higher fidelity and exhibit lower source-modeling systematics \citep{Ephremidze2025}, thereby improving the reliability of dark matter constraints inferred from the residuals. The trade-off is a substantially higher computational cost that will require careful balancing against inference efficiency.

The most pressing theoretical need is for wave-mechanics simulations of $\psi$DM in cluster-mass halos.
The fluctuation amplitude calibration of \citealt{Laroche2022} was derived for galaxy-scale halos and is extrapolated here to cluster environments, introducing an uncertainty that is hard to quantify without dedicated numerical input.
Cluster-scale $\psi$DM simulations would capture physical effects absent from the current model:  soliton core oscillations, coherence-length-set spatial correlations among granules, tidal disruption near the cluster center, and baryonic back-reaction -- and would ultimately replace the free parameter $A_\psi$ with a simulation-based prediction.

Finally, applying the analysis to multiple cluster lenses would reduce source-morphology variance by a factor of $1/\sqrt{N_\mathrm{clusters}}$, thereby extending sensitivity to lower amplitudes and lighter boson masses.
Ongoing JWST cluster programs -- including UNCOVER \citep{Bezanson2022}, MAGNIF \citep{Sun2023}, and Director's Discretionary programs targeting massive lensing clusters -- provide ideal datasets for such an extension. Additionally, combining $P_\delta(k)$ constraints from clusters with those from quad-lens flux ratios \citep{Laroche2022} and VLBI lensing \citep{Powell2023} into a joint inference framework would leverage the complementary environmental and scale dependences of each probe to break degeneracies among $m_\psi$, $A_\psi$, and the halo mass function suppression.

\section{Conclusions}
\label{sec:conclusions}

We have demonstrated a statistical method to probe wave-like dark matter ($\psi$DM) using JWST observations of strongly lensed galaxies in galaxy clusters.
The method is based on the residual power spectrum $P_\delta(k)$ -- the two-dimensional power spectrum of normalized fitting residuals from a smooth lens\,+\,source model -- which encodes the population-level lensing imprint of dark matter substructure without requiring the resolution of individual subhalos.
Our main conclusions are:

\begin{enumerate}
    \item \textit{The residual power spectrum is sensitive to both the boson mass and the fluctuation amplitude of $\psi$DM.}
    
    At the physical wave-interference amplitude $\log_{10}(A_\psi) = -0.8$ from \citet{Laroche2022} and the heavier particle masses ($m_\psi \gtrsim 10^{-21}\,\mathrm{eV}$), the $\psi$DM substructure is CDM-like and the spectra are indistinguishable. As $m_\psi$ becomes lighter, two competing mechanisms control $P_\delta(k)$: the subhalo mass function is increasingly suppressed, while both the scale and amplitude of wave-interference fluctuations grow. At $m_\psi = 10^{-22}\,\mathrm{eV}$, suppression wins: the fluctuations are still too small-scale and low-amplitude to compensate for the loss of subhalos, and $P_\delta(k)$ falls below the CDM prediction. At $m_\psi = 10^{-23}\,\mathrm{eV}$, the larger de Broglie wavelength and stronger fluctuations drive $P_\delta(k)$ above CDM.
    
    At a fixed boson mass, boosting the fluctuation amplitude leads to a monotonic rise in $P_\delta(k)$ while preserving its spectral shape.
    
    \item \textit{The $\psi$DM signal -- where detectable -- manifests as a broadband power-law modification, not a sharp spectral peak.} 
    
    Although the $\psi$DM density fluctuations have a characteristic three-dimensional scale set by $\lambda_\mathrm{dB}$, the two-dimensional projection involved in lensing smears any feature at $k \sim 2\pi/\lambda_\mathrm{dB}$ into a broadband modification of $P_\delta(k)$ that can appear as either an excess (large $\lambda_\mathrm{dB}$ or large $A_\psi$) or a deficit (small $\lambda_\mathrm{dB}$ and low $A_\psi$) relative to CDM. A robust search must therefore compare the measured spectrum against forward-modeled predictions over the full accessible $k$ range, rather than looking for a localized spectral feature.
    
    \item \textit{Realistic strong lensing analysis with CAB lens and shapelet source modeling preserves the $\psi$DM signal.}
    
    The CAB\,+\,shapelet model fit yields stable, well-constrained posteriors regardless of the underlying DM model, reliably capturing the smooth component of the lensing without absorbing the small-scale substructure signal. At our 20-hour JWST depth, however, the post-fit CDM residuals are dominated by source\,+\,lens modeling systematics rather than by the bare CDM subhalo signal (Appendix~\ref{app:systematics}). The $\psi$DM-vs-CDM separations we report should therefore be read as separations between the $\psi$DM signal and a CDM\,+\,systematics baseline.

    \item \textit{A single 20-hour JWST cluster observation can discriminate $\psi$DM with $m_\psi \lesssim 10^{-22}\,\mathrm{eV}$ from CDM.}
    
    At the physical amplitude, $m_\psi = 10^{-23}\,\mathrm{eV}$ is cleanly separated from the CDM\,+\,systematics baseline, with non-overlapping 16th--84th percentile bands across the full analysis range $1 \lesssim k \lesssim 11\,\mathrm{kpc}^{-1}$. For $m_\psi = 10^{-22}\,\mathrm{eV}$ at $\log_{10}(A_\psi) = 1.2$ ($100\times$ the physical prediction), the $\psi$DM spectrum sits a factor of $\sim\!3$--$10$ above the baseline across $k \sim 2$--$11\,\mathrm{kpc}^{-1}$, again with non-overlapping bands. More advantageous lens configurations (e.g. a galaxy lens in a cluster lens) and improved modeling methods with lower systematics may allow us to probe $\psi$DM with higher particle masses. 
\end{enumerate}

Taken together, these results establish the residual power spectrum as a promising window into the wave-like nature of dark matter, offering a population-level statistical probe that is independent of, and complementary to, existing constraints from individual lens systems.

\section*{Acknowledgements}
We would like to acknowledge useful discussions with Atınç Çağan Şengül, Chandrika Chandrashekar, and Hadrien Paugnat. CD and NE were partially supported by the Department of Energy (DOE) Grant No. DE-SC0025671. DG acknowledges support from the Brinson Foundation provided through a Brinson Prize Fellowship Grant. 

\section*{Data Availability}

The code used to generate the mock images, perform the lens modeling, and compute the residual power spectra is publicly available at \url{https://github.com/ninoephremidze/statistically_probing_subgalactic_dark_matter}.

\appendix

\section{Modeling-systematics in $P_\delta(k)$}
\label{app:systematics}

A central question raised by the CDM residuals in \S\ref{sec:results_reconstructions} is whether the small, coherent residual structure visible there originates from genuine lensing perturbations by CDM subhalos or from imperfections in our source and lens modeling. The per-pixel lensing perturbation from subhalos in our rendered range ($10^6 \le M_\mathrm{sub}/M_\odot \le 10^9$) is small, and the CDM ensemble signal could plausibly sit at, near, or below the modeling-systematics floor at 20-hour JWST depth.

\paragraph*{Null test on smooth mocks.}
To isolate this floor we generate 20 realizations of \emph{smooth} mocks (60 residual maps in total across the three images per realization), in which no substructure of any kind is rendered, but every other ingredient of the pipeline described in \S\ref{sec:methods} is retained. Any non-zero $P_\delta(k)$ measured from these mocks is by construction attributable to modeling systematics rather than to dark-matter substructure.

\begin{figure}
    \centering
    \includegraphics[width=\linewidth]{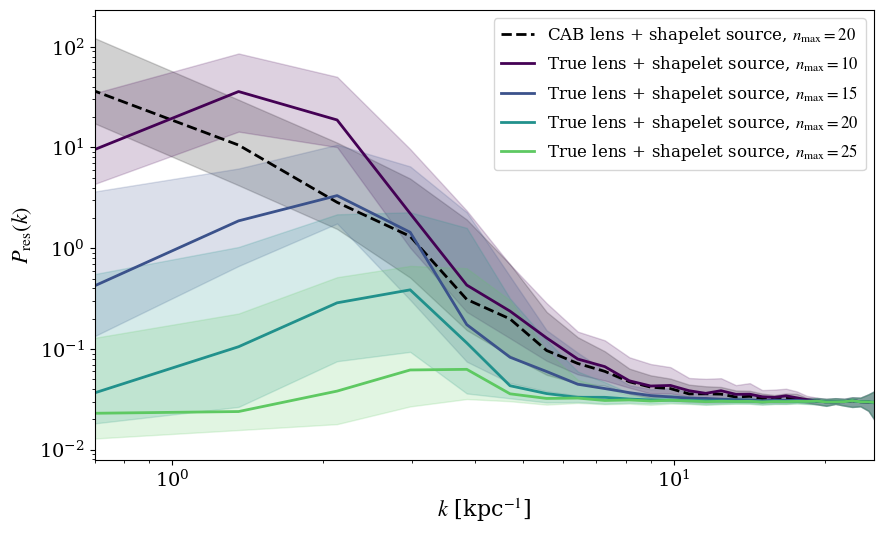}
    \caption{Residual power spectrum $P_\delta(k)$ of 60 smooth (no-substructure) mocks at $t_\mathrm{exp}=20\,\mathrm{h}$.
    \textit{Black}: joint CAB\,+\,shapelet fit using $n_\mathrm{max}=20$ -- the full pipeline of the main analysis.
    \textit{Coloured}: source-modeling-only null test, in which the lens is fixed to the true local elliptical-NFW deflection field used to generate the mocks and only the shapelet amplitudes are solved for. Shaded bands show the 16th--84th percentile across the 60 maps. The gap between the black and the teal curves (both at $n_\mathrm{max} = 20$) at a given $k$ is the contribution of lens modeling (CAB-vs-true-NFW approximation error) to the joint-fit floor at that scale: where the two coincide the floor is source-modeling-limited; where the coloured curve sits below the black curve the floor is lens-modeling-limited.}
    \label{fig:systematics}
\end{figure}

We then fit the smooth mocks with the same CAB\,+\,shapelet pipeline used in the main analysis and compute $P_\delta(k)$ on the resulting residual maps. The black dashed curve in Figure~\ref{fig:systematics} shows this \emph{joint-fit systematics floor}: the residual power that the full pipeline produces even when there is no substructure signal to recover.

\paragraph*{Decomposing source vs.\ lens model systematics.}
To split this floor into its source- and lens-modeling components, we run a second set of fits on the same 60 smooth mocks in which the lens model is fixed to the true local elliptical-NFW deflection field that generated the mocks and only the source-side shapelet amplitudes are solved for. Any residual power in this null configuration is, by construction, attributable to source-modeling error alone -- specifically, the failure of a given shapelet basis to represent the morphological complexity of COSMOS sources at the noise level. We repeat this source-only fit at several shapelet expansion orders $n_\mathrm{max}$; the coloured curves in Figure~\ref{fig:systematics} show the resulting $P_\delta(k)$ at each $n_\mathrm{max}$, illustrating how the source-only floor depends on the flexibility of the shapelet basis.

The source- vs.\ lens-modeling decomposition is meaningful only between the joint-fit curve and the source-only curve evaluated at the \emph{same} shapelet order. Throughout the main analysis we use $n_\mathrm{max}=20$, so the relevant pair is the joint-fit curve (black dashed in Figure~\ref{fig:systematics}) and the source-only curve (teal) both at $n_\mathrm{max}=20$. These two admit a three-regime interpretation:
\begin{itemize}
    \item At the highest wavenumbers $k$, where both curves coincide with the white-noise plateau, the shapelets are not fitting noise and the floor is set by the per-pixel noise itself.
    \item At intermediate scales ($k \sim 3$ kpc$^{-1}$), where the two curves almost coincide above the white-noise plateau, the source modeling systematics contribute significantly to the joint-fit residual.
    \item At larger scales, where the source-only curve sits below the joint-fit curve, the gap corresponds to the lens-modeling contribution (the residual error from approximating the true NFW deflection field by the local CAB expansion) and dominates the joint-fit systematics floor.
\end{itemize}
The other coloured curves show that the source-only floor decreases monotonically with $n_\mathrm{max}$, indicating that an increasingly complex source model reduces the systematics, as expected.

\paragraph*{Implications for the main results.}
The black curve in Figure~\ref{fig:systematics} lies at essentially the same amplitude as the post-fit CDM $P_\delta(k)$ shown in Figures~\ref{fig:ps_case1} and~\ref{fig:ps_case2}. We therefore conclude that the post-fit CDM curve in our main results is dominated by these modeling systematics rather than by the bare lensing signal from CDM subhalos in the rendered mass range. Two consequences follow.

First, the $\psi$DM-vs-CDM separations reported in \S\ref{sec:results_ps} should be interpreted as separations between the $\psi$DM signal and a CDM\,+\,systematics floor: this is the relevant baseline for distinguishing $\psi$DM from a no-detection scenario in real data, but it is not a direct comparison between bare CDM and bare $\psi$DM lensing signals at this exposure depth.

Second, the systematics floor sets the sensitivity limit of the present analysis. Lowering it would require improved, more complex and flexible source and lens modeling techniques, discussed in more detail in \S\ref{sec:future}.

\section{Posterior Distributions of CAB Parameters}
\label{app:posteriors}

\begin{figure*}
    \centering
    \includegraphics[width=\linewidth]{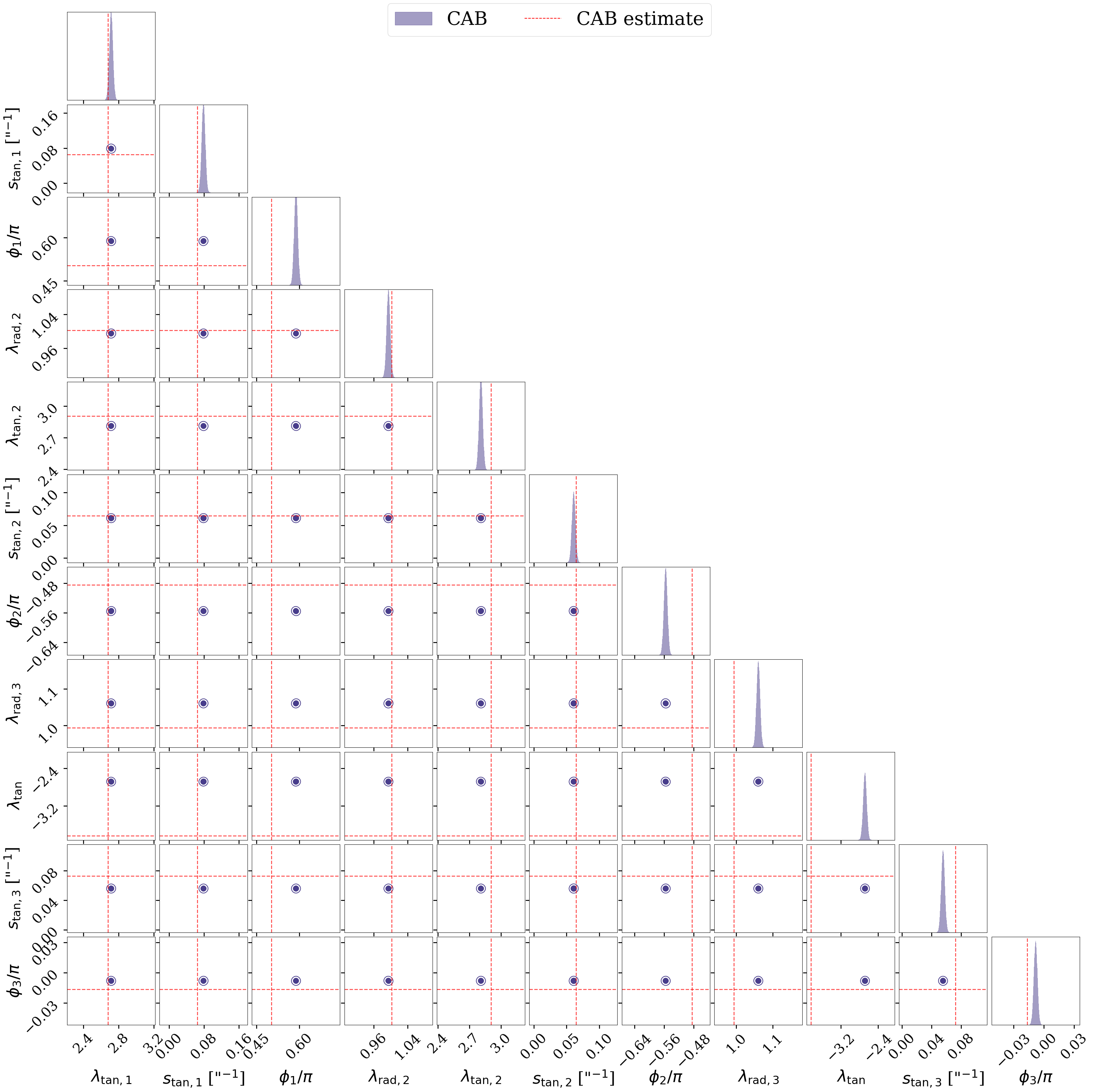}
    \caption{Marginalized posterior distributions of the CAB lens parameters recovered by the \textsc{dynesty} nested sampler for a representative \textbf{CDM} mock realization. Red lines indicate the CAB estimate computed from the NFW macro model. All posteriors are well-constrained, with no significant inter-parameter correlations, confirming that the smooth CDM signal is well described by the CAB model.}
    \label{fig:posteriors_cdm}
\end{figure*}

\begin{figure*}
    \centering
    \includegraphics[width=\linewidth]{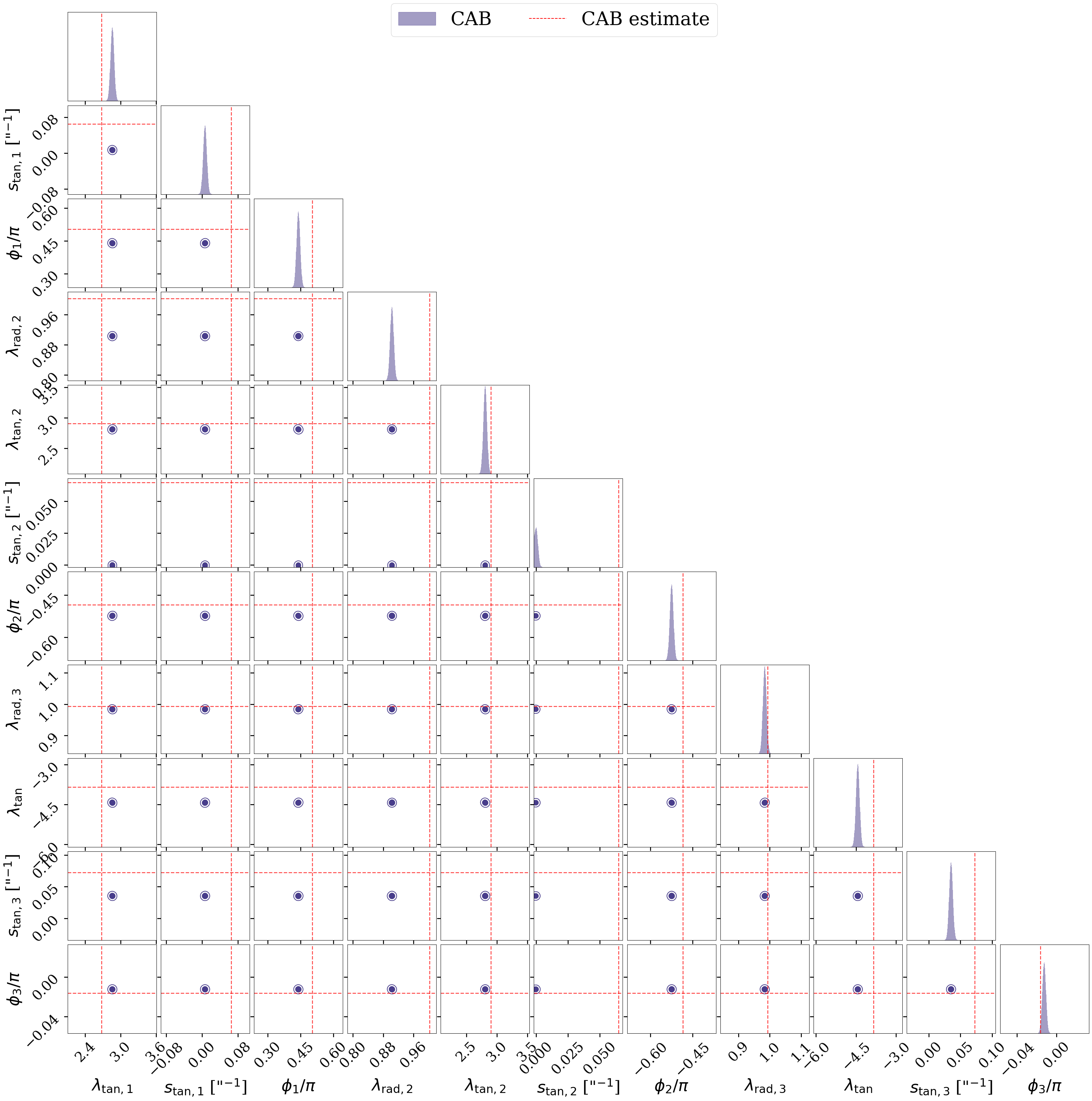}
    \caption{Same as Figure~\ref{fig:posteriors_cdm}, but for a \textbf{Case~1} $\psi$DM realization ($m_\psi = 10^{-22}\,\mathrm{eV}$, $100\times$ the physical amplitude). The posteriors are similarly well constrained as in the CDM case, demonstrating that the strong $\psi$DM density fluctuations are not absorbed into the smooth lens parameters and instead produce the coherent residual structure quantified by $P_\delta(k)$.}
    \label{fig:posteriors_case1}
\end{figure*}

\begin{figure*}
    \centering
    \includegraphics[width=\linewidth]{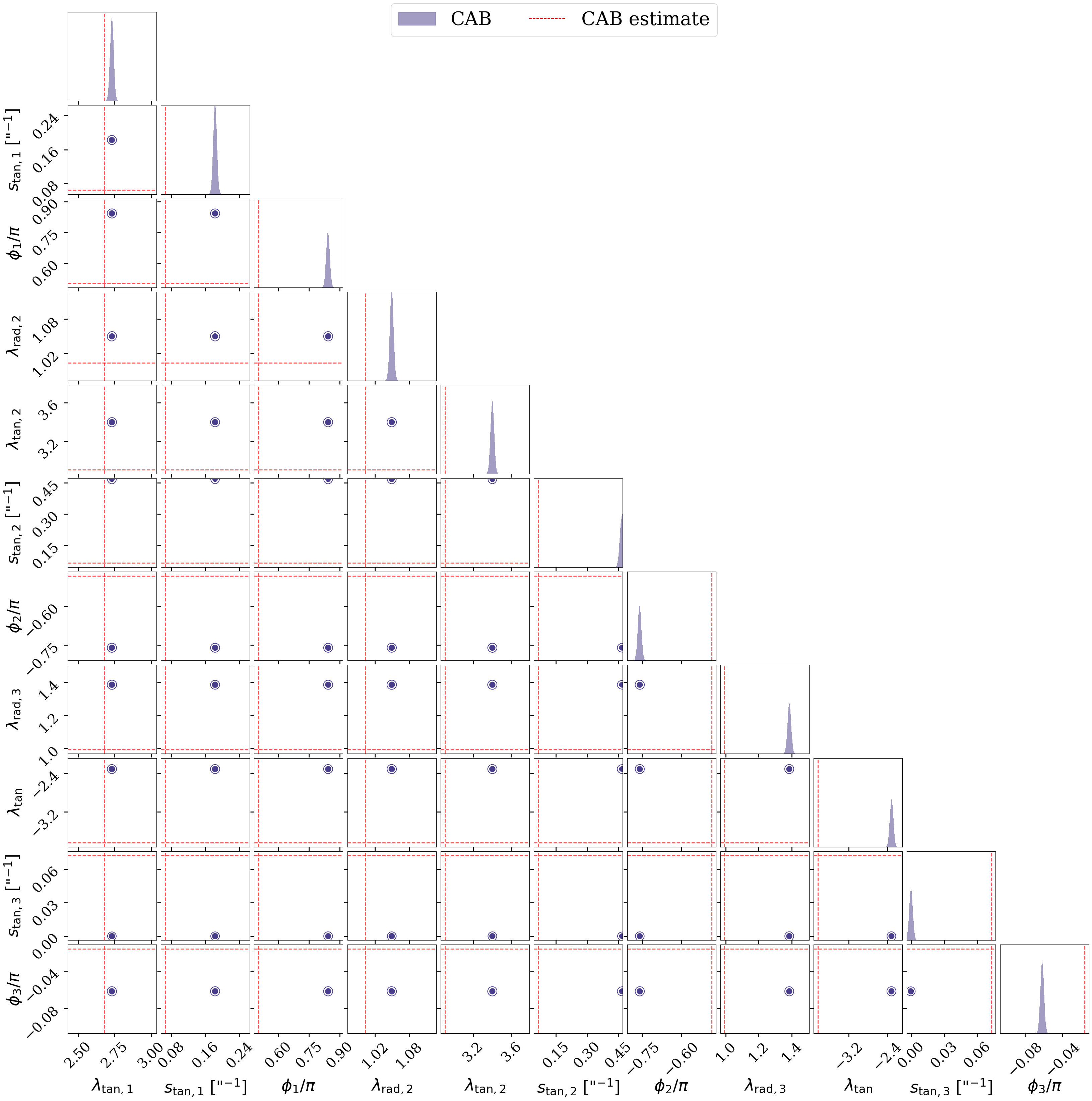}
    \caption{As Figure~\ref{fig:posteriors_cdm}, but for a \textbf{Case~2} $\psi$DM realization ($m_\psi = 10^{-23}\,\mathrm{eV}$, $\log_{10}A_\psi = -0.8$, the physical prediction of \citealt{Laroche2022}). The posteriors remain well constrained but deviate further away from the CAB estimate, reflecting a mild degeneracy between the large-scale $\psi$DM fluctuations at this low boson mass and the smooth lens parameters.}
    \label{fig:posteriors_case2}
\end{figure*}

We examine the posterior distributions of the 11 free CAB lens parameters
($\lambda_\mathrm{tan}$, $s_\mathrm{tan}$, $\phi$ for each of the three lensed images, plus $\lambda_\mathrm{rad}$ for images~2 and~3, with $\lambda_\mathrm{rad}$ of image~1 fixed to unity)
recovered by the nested sampler for each dark matter model.
The 4 positional-shift components and 3 source parameters are omitted from the corner plots for clarity, as they are nuisance parameters with no sensitivity to the dark matter model.
Figure~\ref{fig:posteriors_cdm} shows the marginalized posteriors for a representative CDM realization,
and Figures~\ref{fig:posteriors_case1}--\ref{fig:posteriors_case2} show the corresponding results for the two $\psi$DM cases.

In all three cases, the posteriors are well-constrained: the marginal distributions are narrow, nearly Gaussian peaks that are mostly to the CAB estimates (red lines; derived from smooth elliptical NFW model without substructure). Any deviation between the posteriors and estimates indicate that some of the large-scale perturbation from substructure is absorbed by the local CAB lens model.

\bibliographystyle{mnras}
\bibliography{main} 
\label{lastpage}
\end{document}